# Near-mid infrared spectroscopy of carbonaceous chondrites: Insights into spectral variation due to aqueous alteration and thermal metamorphism in asteroids

Authors:

Jinfei Yu [1,2,3]

Haibin Zhao (meteorzh@pmo.ac.cn) [1,2,4*]

Edward A. Cloutis [5]

Hiroyuki Kurokawa [3,6]

Yunzhao Wu [1,7]

[1] Key Laboratory of Planetary Sciences, Purple Mountain Observatory, CAS, Nanjing 210023, China

[2] School of Astronomy and Space Science, University of Science and Technology of China, Hefei 230026, China

[3] Department of Earth Science and Astronomy, Graduate School of Arts and Sciences, The University of Tokyo, 3-8-1 Komaba, Meguro-ku, Tokyo 153-8902, Japan

[4] CAS Center for Excellence in Comparative Planetology, CAS, Hefei 230026, China

[5] Centre for Terrestrial and Planetary Exploration, University of Winnipeg, Winnipeg, Manitoba R3B 2E9, Canada

[6] Department of Earth and Planetary Science, Graduate School of Science, The University of Tokyo, 7-3-1 Hongo, Bunkyo-ku, Tokyo 113-0033, Japan

[7] State Key Laboratory of Lunar and Planetary Sciences, Macau University of Science and Technology, Macau, China

## Abstract:

Carbonaceous chondrites (CCs) are windows into the early Solar System and the histories of their parent bodies. Their infrared spectral signatures are powerful proxies for deciphering their composition and evolution history, but still present formidable challenges in relation to determining the degree of secondary processing such as aqueous alteration and thermal metamorphism via comprehensive data and mid-infrared feature. In our study, we delved into the infrared spectra spanning 1–25 μm of 17 CCs, with distinct petrological characteristics and varying degrees of alteration. Through this investigation, we uncovered distinct spectral patterns that shed light on the processes of alteration and metamorphism. As aqueous alteration intensifies, two key spectral features, the 3 μm-region absorption feature associated with OH-bearing minerals and water, and the 6 μm band indicative of water molecules, both grow in intensity. Simultaneously, their band centers shift towards shorter wavelengths. Moreover, as alteration progresses, a distinctive absorption feature emerges near 2.72 μm, resembling the OH absorption feature found in serpentine and saponite minerals. Comparison of aqueous alteration to laboratory-heated CCs suggests that the 3 μm region OH/$H_2O$ absorption feature differs between CC heated to less than or more than ~300°C. Further insights are gained by examining the vibrational features of silicate minerals, notably influencing the 10 μm and 20 μm regions. The 12.4 μm /11.4 μm reflectance ratio diminishes, and the reflectance peak in the 9–14 μm range shifts towards shorter wavelengths. These changes are attributed to the transformation of anhydrous silicates into phyllosilicates. In the 15–25 μm region, the influence of thermal metamorphism becomes evident and results in the appearance of more spectral features, the single reflectance peak at 22.1 μm undergoes a transformation into two distinct peaks at 19 μm and 25 μm, which is primarily attributed to the increased presence of anhydrous silicates and olivine recrystallization. These findings offer novel insights into the volatile-rich compositions of C-complex asteroids and the thermal evolution histories of their parent bodies.

## 1. Introduction

Small Solar System bodies offer a unique opportunity to delve into the evolutionary processes that have shaped our Solar System (McSween and Lauretta, 2006). Among these objects, C-complex asteroids (C-, D-, G-, F-, and B-asteroids) and carbonaceous chondrites (CCs) have long been considered closely related, primarily due to the striking similarities observed in their spectral characteristics (McSween and Lauretta, 2006; Bottke et al., 2015). These shared features include similar spectral slopes, low albedo, and a prominent 3 μm region absorption band, which is a characteristic of hydrous materials (Binzel et al., 1989; Bühler, 1988). These objects preserve a record of aqueous alteration active in young planetesimals, thus providing an opportunity to understand the distribution and evolution of volatile components in the early Solar System, and to understand water-rock reactions in those planetesimals. Furthermore, some asteroids and CCs appear to have undergone dehydration processes induced by solar radiation, radiogenic decay, vacuum desiccation, or impact-related heating (Hiroi et al., 1993, 1996b; Beck et al., 2018; King, 2021; Matsuoka et al., 2022). This inference is drawn from their distinct spectral signatures, which deviate slightly from those of pristine or space-weathered chondrites (Matsuoka et al., 2015, 2020). Instead, they exhibit spectral characteristics akin to experimentally-heated chondrites, so they retain a record of thermal metamorphism events: as repositories of crucial information about their subsequent evolution after early stage alteration (Hiroi et al., 1993, 1996b).

Laboratory analyses of CCs combined with observations of C-complex asteroids have the potential to set essential constraints on the mechanism, time, and place of both processes, providing a novel window into the effects of asteroidal processing in the early Solar System. Previous studies have extensively examined the intricate mechanisms of aqueous alteration processes through meticulous petrological and geochemical analyses, elucidating that aqueous alteration involves the low-temperature interaction between fluids and rocks, which resulted in in the transformation of anhydrous silicates into phyllosilicates and the gradual transition from Fe-rich phases to Mg-rich phases (Browning et al., 1996; Howard et al., 2011, 2009; McSween, 1987; Rubin et al., 2007; Zolensky et al., 1993).

Infrared spectroscopy serves as a vital tool for detecting hydrated minerals (Beck et al., 2014; Milliken and Mustard, 2007a; Miyamoto and Zolensky, 1994; Hiroi et al., 1996b), especially in the 3 μm region. An absorption feature in this region is related to the presence of OH/$H_2O$ in a variety of materials (Miyamoto and Zolensky, 1994; Takir et al., 2013; Garenne et al., 2014; Beck et al., 2010, 2018). It remains the primary technique to unravel the extent of aqueous alteration and thermal metamorphism in volatile-rich asteroids, and many questions such as how, where, and when this aqueous alteration occurred can be constrained, offering a unique glimpse into the effects of processing on early Solar System materials. The JAXA mission Hayabusa2 (Watanabe et al., 2017), and the NASA mission Origins Spectral Interpretation Resource Identification Security-Regolith Explorer (OSIRIS-REx; Lauretta et al., 2017) investigated C-complex asteroids Cg-type asteroid Ryugu and B-type asteroid Bennu respectively. Analysis of the global near-infrared (NIR) spectrum of these near-Earth objects show distinct spectral features: Ryugu's 3 μm-region absorption feature is weak and sharp, similar to the CY and heated CI1 chondrites (Kitazato et al., 2019), while Bennu's absorption feature is deeper and intermediate between the CI1 and CM2 chondrites (Hamilton et al., 2019), leading to the suggestion that materials on Ryugu's surface experienced either a low degree of alteration or dehydration following alteration compared to the more highly-altered Bennu (Galiano et al., 2020; Kitazato et al., 2021). However, it is worth noting that returned samples from Ryugu unexpectedly appear to lack the thermal metamorphism and consequent dehydration observed in previous studies (E. Nakamura et al., 2022; T. Nakamura et al., 2022; Yokoyama et al., 2022). In addition, the NASA Dawn mission (Russell and Raymond, 2011), the NASA Infrared Telescope Facility (Takir and Emery, 2012) and the JAXA AKARI infrared space telescope (Usui et al., 2019) obtained more reflectance spectra of large main belt asteroids and revealed the complexity and uniqueness of the 3 μm band, such as -OH absorption feature around ~2.75 μm akin to meteorites (Potin et al., 2020), and a ~3.1 μm absorption feature, attributed to the presence of water ice, ammoniated phyllosilicates, or goethite (Campins et al., 2010; De Sanctis et al., 2015; Beck et al., 2011), which suggest that the process of aqueous alteration of outer main belt objects is controlled by other factors such as volatile components (Kurokawa et al., 2022).

However, the distribution of asteroids' 3 μm-region absorption feature and how it relates to the history of those asteroids' evolution are still contentious issues. On one hand, we still lack more 3 μm region data comparisons of asteroids with meteorites, on the other hand, we lack samples from asteroids. In parallel, the presence of a 0.7 μm absorption feature has been reported for some asteroids from both telescopic observation and asteroid exploration missions (Hiroi et al., 1993; Kameda et al., 2021; McAdam et al., 2015). Also, due to the weakness of this absorption feature, the relationship with alteration and metamorphism remains unclear. When we explore the mid-infrared (MIR) region, we observe key bands: the 6 μm band is associated with the bending vibrations of water molecules (Honniball, 2021) and the 9–14 μm region and 15–25 μm regions are indicative of silicate vibrations (Bates, 2020; Bates et al., 2020a; Donaldson Hanna et al., 2019). However, the connections between these MIR features and processes like aqueous alteration and thermal metamorphism are still less understood, requiring ongoing research.

To gain a comprehensive understanding of the relationship between the aforementioned near-to-mid-infrared (NIR-MIR) features of CCs and the aqueous alteration and thermal metamorphism processes on their parent bodies, this study meticulously analyzes NIR-MIR spectral data from 17 CCs with varying degrees of alteration, alongside data from controlled heating experiments on one CC sample. Our primary aim is to unravel the spectral variations and factors controlling aqueous alteration and thermal metamorphism in CCs, with specific focus on the spectral characteristics of hydrated minerals, especially phyllosilicates. Furthermore, we compare our findings with previously published petrological studies. We also delve into the application of the spectral relationships of CCs to asteroid spectral data and their integration with petrological information. This comprehensive analysis aims to provide support to elucidate the compositional and evolution trends of surface materials on C-complex asteroids with future infrared observation.

# 2. Spectral data and methods

## 2.1. Sample Selection

This study focuses on the spectra of 17 CCs with different petrological type and one CC heating experiment (Table 1). The chosen samples are from various groups and petrological types, representing CCs that have undergone various degrees of aqueous alteration and thermal metamorphism as described by several authors (Howard et al., 2015; Huss et al., 2006; Rubin et al., 2007). Specifically, the sample set includes 7 CCs from the CM group, 3 CCs from the CI group, 1 CC from the CR group, 1 CC from the CO group, 1 CC from the CV group, 3 CCs from the CK group, and 1 ungrouped C2 chondrite. The petrological types range from type 1 to type 4 (4/5) (see Table 1). For CC heating experiment, bulk samples of the Murchison meteorite were placed in a moderate vacuum environment (hydrogen, $10^{-5}$ atm) and heated continuously for 1 week at temperatures of 400°C, 500°C, 600°C, 700°C, 800°C, 900°C, and 1000°C. Reflectance spectra were obtained from the NASA Reflectance Experiment Laboratory (RELAB) at Brown University (Milliken et al., 2016), The spectroscopic measurements were conducted using a UV-VIS-NIR bidirectional reflectance spectrometer (0.3–2.6 μm) and a Thermo Nexus 870 FT-IR spectrometer (1.4–25 μm). Reflectance spectroscopy measures the reflected light intensity of powdered samples, providing a function of wavelength versus brightness. This method is rapid and requires simple sample preparation. To minimize the influence of geometric shape and particle size on the measurements, all meteorite samples were ground and sieved to a grain size smaller than 125 μm.

## 2.2. Infrared spectroscopy

This study focuses on the spectral variations in the NIR region (1–5 μm) and MIR region (5–25 μm) of CCs during aqueous alteration and thermal metamorphism processes. Due to the generally weak absorption features in the infrared spectra of CCs, it is necessary to perform continuum removal to effectively enhance the absorption features for accurate identification of key absorption peaks and absorption bands measurement of their positions and depths or intensities. In this study, the ENVI software was used for continuum removal of the sample spectral data. The principle of continuum removal involves fitting an envelope curve (continuum) to the spectral curve. This envelope curve is formed by connecting the peak points on the spectral and fitting them with a polynomial function. It creates a convex hull curve above the spectral curve connecting local maxima with straight line segments. The depth of an absorption peak in the spectrum, *Db*, is given by,

$$D_{\mathrm{b}} = \frac{R_{\mathrm{c}} - R_{\mathrm{b}}}{R_{\mathrm{c}}} \ . \quad (1)$$

Here, *Rb* is the reflectance of the spectrum after continuum removal, and *Rc* is the reflectance of the continuum at the same wavelength as Rb. In this study, the absorption intensity and absorption center position of the spectral curve at the 3 μm and 6 μm bands were measured. For the continiuum removal starting and ending point wavelengths are 2.6–4.0 μm (3 μm band ) and 5.6–6.60 μm (6 μm band). The mineral compositions of CCs are also identified by referring to the mineral spectral data from NASA's Planetary Data System (PDS) and other infrared spectroscopy studies (Takir et al., 2013; Beck et al., 2010, 2018; Bates et al., 2020b, 2020a; Potin et al., 2020). Furthermore, the differences in the spectral shapes between CCs with varying degrees of aqueous alteration and metamorphism were compared in the 9–14 μm and 15–25 μm regions. The reflectance ratio between 11.4 μm and 12.4 μm was calculated, and the analysis was combined with the petrologic characteristics of the samples.

**Table 1**

List of samples in this study. Bulk H, phyllosilicate, and water content for CI and CM meteorites are referenced from previous studies (Alexander et al., 2012; King et al., 2015, 2017; Howard et al., 2009, 2011; Braukmüller et al., 2018). Note that the samples are still affected by atmospheric alteration during preservation and measurement, resulting in potentially inaccurate water content. Actual water content is invariably lower than the values reported here.

| Name | Classification | Fall or Find | Wavelength range /μm | Refl. 2 μm | Total Phyllosilicate (vol%) | H Bulk (wt%) | Water (wt%) |
|---|---|---|---|---|---|---|---|
| Alais | CI1 | Fall, 1806 | 0.83–99.72 | 0.07370 | 83.0 | - | - |
| Orgueil | CI1 | Fall, 1864 | 1.43–25.05 | 0.06885 | 84.0 | 1.561 | 18.14 |
| Ivuna | CI1 | Fall, 1938 | 0.83–99.72 | 0.04305 | 84.0 | 1.524 | - |
| Moapa Valley | CM1 | Find, 2004 | 0.83–99.72 | 0.03308 | 87.5 | - | - |
| QUE97077 | CM2 (2.6) | Find, 1997 | 0.32–25.05 | 0.07143 | - | - | - |
| Murchison | CM2 (2.5) | Fall, 1969 | 0.32–25.05 | 0.05327 | 72.5 | 0.96 | 10.36 |
| Murray | CM2 (2.4) | Fall, 1950 | 0.32–25.05 | 0.10355 | 74.0 | 1.03 | 11.64 |
| Nogoya | CM2 (2.2) | Fall, 1879 | 0.32–25.05 | 0.06777 | 75.8 | 1.31 | 14.72 |
| Mighei | CM2 | Fall, 1889 | 0.32–25.05 | 0.06453 | 74.6 | 0.99 | 10.28 |
| Cold Bokkeveld | CM2 (2.2) | Fall, 1838 | 0.32–25.05 | 0.06267 | 77.4 | 1.22 | 15.61 |
| Al Rais | CR2 | Fall, 1957 | 0.83–99.72 | 0.04926 | 60.0 | 0.893 | - |
| ALHA77307 | CO3 (3.0) | Find, 1977 | 0.83–99.72 | 0.05658 | - | - | - |
| Allende | CV3 (3.0) | Fall, 1969 | 0.83–99.72 | 0.07053 | - | - | 0.08 |
| EET92002 | CK4 | Find, 1992 | 0.90–24.92 | 0.03001 | - | - | - |
| Y-693 | CK4 | Find, 1969 | 1.80–25.99 | 0.05464 | - | - | - |
| EET87507 | CK4/5 | Find, 1987 | 1.80–25.99 | 0.07179 | - | - | - |
| Tagish Lake | C2ung | Fall, 2000 | 1.43–25.05 | 0.02612 | - | 0.872 | 11.62 |

# 3. Results

The spectra obtained are presented in two parts: the 0.35–4.2 μm range and the 5–25 μm range. For the study of aqueous alteration processes, CCs samples representing various degrees of aqueous alteration from the CI, CM, CO, CR groups, and CV Allende, were selected for analysis. Additionally, the thermally metamorphosed CK sample EET92002 were chosen for comparison. For thermal metamorphism study, laboratory heated CM

Murchison, all CK group samples, unheated CM Murchison and unheated CI Ivuna was selected for comparison. Spectral data for all samples can be found in Figure S1 of the supplementary material.

### 3.1. 3 μm region absorption feature

A 3 μm-region absorption feature is the most significant spectral feature of hydrated CCs. It consists of two types of absorption features. One is associated with structural OH present in phyllosilicates and is due to a  stretching vibration, with different mineral structures exhibiting absorption center position typically within the range of 2.7–2.85 μm (Bishop et al., 2008). The absorption feature for iron-rich phyllosilicates is located around 2.85 μm (Bühler, 1988), while the absorption feature of magnesium-rich phyllosilicates is located around 2.72 μm (Milliken and Mustard, 2007a). The $H_2O$ asymmetric and symmetric stretching vibrations produce a rounded absorption near 2.95 μm (Kaplan et al., 2019). The $H_2O$ giving rise to this feature can arise from a number of sources of $H_2O$, the primary contribution likely being due to terrestrial water. Takir et al. (2013) and Matsuoka et al. (2022) have noted changes in the 3 μm band by adsorbed water. Beck et al. (2010) found that the wavelength position of the OH absorption band is largely unaffected by the presence of the adjacent $H_2O$ feature. Therefore the minimum of the OH feature can be used as an indicator of the actual position of this feature without continuum removal. In the ensuing discussion we use the term “3 μm” to refer to the OH absorption feature in the 2.7–2.85 μm region.

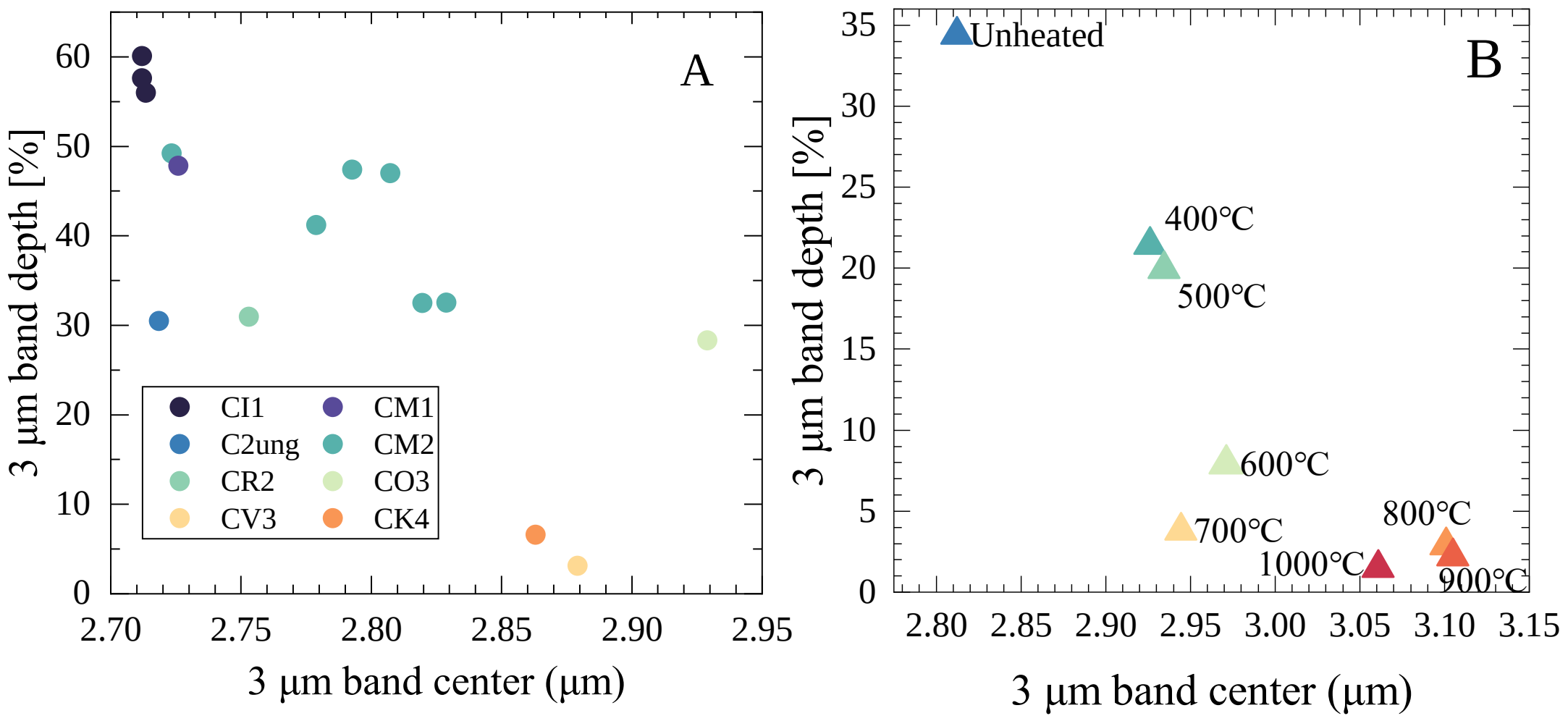

*Figure 1 variation of the 3 μm band center vs. band depth for CC samples. (A) For CC samples with different alteration degree. Note that absorption of CO, CV, and CK meteorites are mainly from terrestrial adsorbed water. (B) For Murchison meteorite subjected to varying temperature heating.*

Figure 1 shows the variation of the 3 μm band depth and position with aqueous alteration degree in the studied CCs. It can be seen that the 3 μm band center shifts to shorter wavelengths as the degree of alteration increases (Figure 1A), moving from ~2.9 μm (ALHA77307, the least altered CO3 meteorite, which has rare phyllosilicate phases that are locally present in the matrix; Brearley, 1993) to 2.72 μm (all CI1 meteorites, the most altered). This variability in the shape of the 3 μm band has been noted previously (Beck et al., 2018; Takir et al., 2013). The leftward shift of the absorption center is interpreted as a variability of phyllosilicates according to the analysis of the mineralogy of these samples (Browning et al., 1996; Howard et al., 2009, 2011).

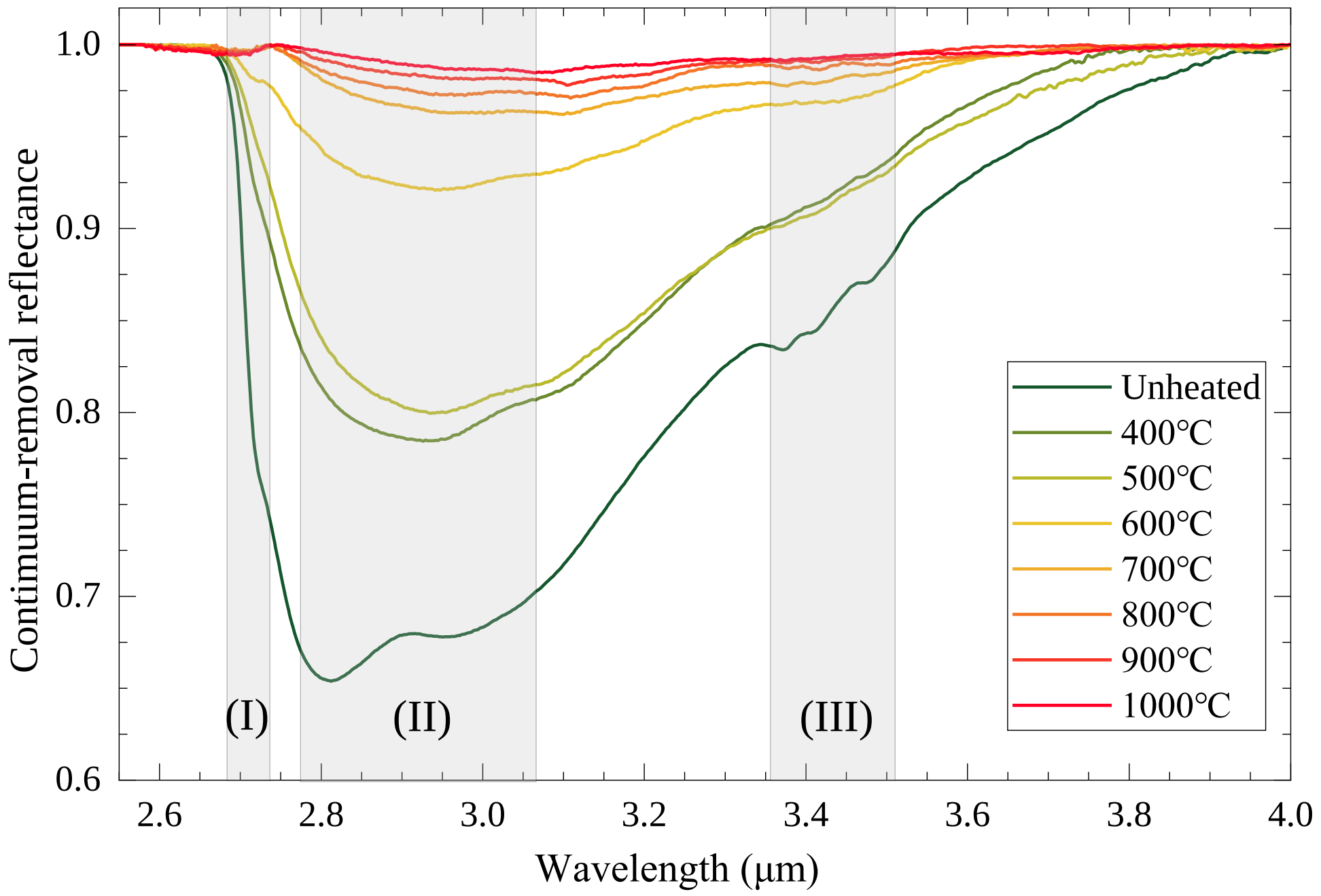

*Figure 2 Continuum-removal spectra of Murchison meteorite heated at different temperatures at 3 μm band. The shaded areas in the figure indicate: (I) Mg-rich serpentine-like minerals, (II) Fe-rich cronstedtite-like minerals and $H_2O$, (III) aliphatic and aromatic organic compounds and carbonates.*

During the heating process, the absorption center gradually shifts towards longer wavelengths (Figure 1B and Figure 2). As Mg-rich serpentine undergoes metamorphism and transforms into dehydrated minerals, it leads to a weakening of the absorption feature in the 2.72–2.77 μm region. In addition, the heating and metamorphism of the phyllosilicate may involve $Si^{4+}$ and $Mg^{2+}$ ionic substitution by $Fe^{3+}$ within the mineral structure (King, 2021; Nakamura, 2005), resulting in a relative increase in the content of Fe-rich cronstedtite (while the total amount of phyllosilicate decreases), which causes the -OH absorption feature to shift towards longer wavelengths. The absorption center at longer wavelengths and broader absorption features also corresponds to the formation of iron-bearing hydrated minerals whose absorption features are around the 2.9–3.1 μm range, such as metamorphosed Fe-rich cronstedtite and ferrihydrite (Milliken and Mustard, 2007a). These combined effects result in the weakening of the absorption band and a shift towards longer wavelengths in the 3 μm band of Murchison meteorite during the heating process. Additionally, the "triplet" absorption features around 3.4 μm disappear after heating at temperatures of 400°C and above, indicating the aromatization, dehydration, and volatilization of organic matter (Duan et al., 2021; Kaplan et al., 2018).

### 3.2. 6 μm band

The 6 μm band is a symmetric single absorption, with the absorption center located at 6.07–6.15 μm. This absorption is attributed to the bending vibrations of $H_2O$ molecules associated with hydrous minerals, interlayer water, pore water, adsorbed water, and ambient water in CCs (Beck et al., 2014; Honniball, 2021). Figure 3 illustrates the relationship between the absorption depth, absorption center, and degree of alteration and metamorphism in the 6 μm band.

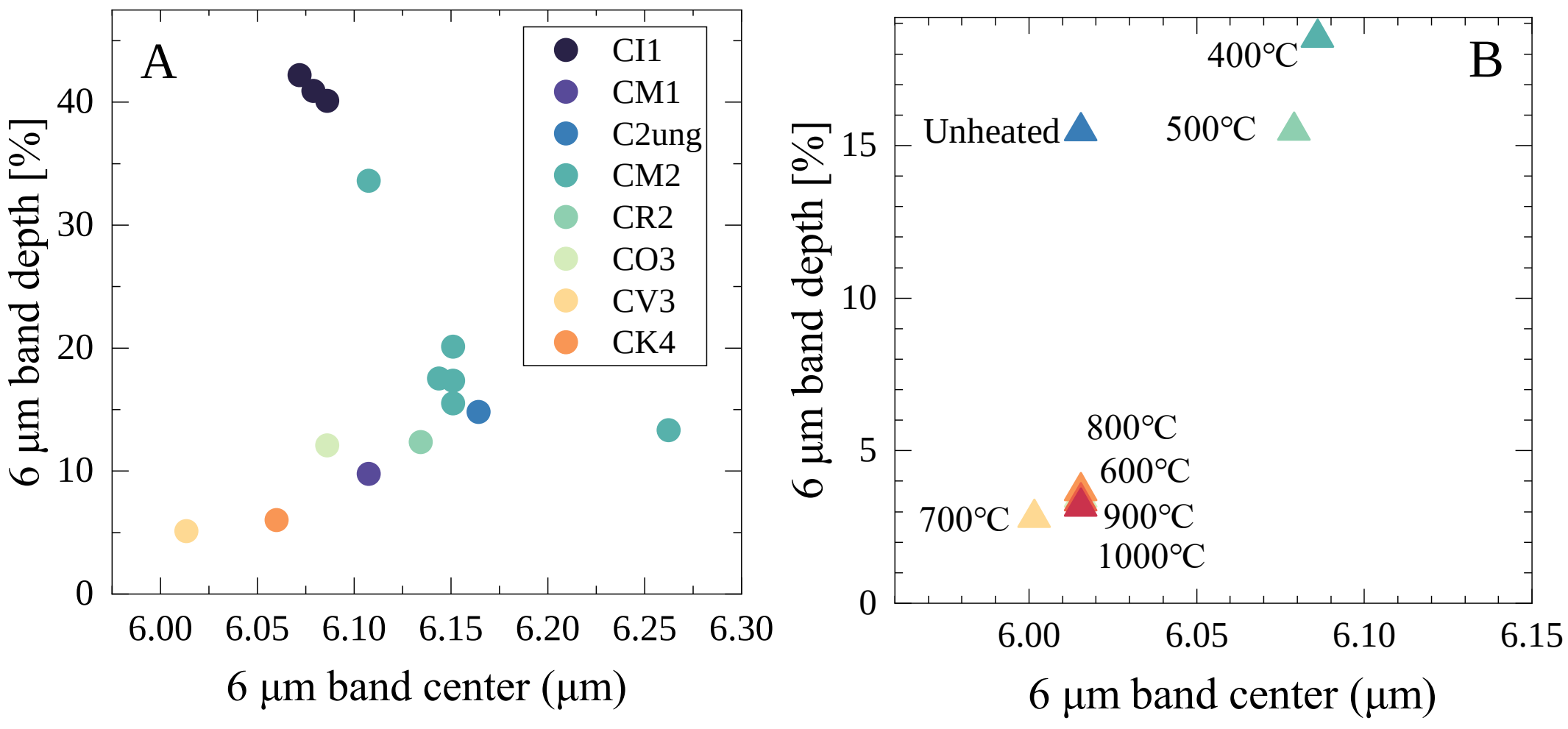

*Figure 3. The variation of the 6 μm band center vs. band depth. (A) For CC samples with different alteration degree. Note that absorption of CO, CV, and CK meteorites are mainly from terrestrial adsorbed water. (B) For Murchison meteorite subjected to varying temperature heating.*

The CM2 group of CCs exhibits significant variations in the absorption depth of the 6 μm band, which is attributed to the varying degrees of alteration in CM2 samples that significantly affect the total water content of the meteorites (Table 1). The CI1 group of meteorites exhibits similar absorption depths and center positions of the 6 μm band. This suggests that they have undergone complete alteration, and the total water content within these meteorites is also similar. The CK4 and CV3 samples have absorption center positions concentrated around 6.03 μm. It significantly differs from CM2 and CI1 meteorites, whose absorption centers are located around 6.09–6.15 μm. The discrepancy in band position is attributed to the different forms of water. Altered meteorites are primarily influenced by mineral water in phyllosilicates: varying mineralogical structures lead to observed differences in band position. In contrast, unaltered meteorites are mainly influenced by terrestrial water in their olivine-dominated (anhydrous silicate) powder, exhibit a weak absorption feature around 6.0 μm (Beck et al., 2018; McAdam et al., 2015).

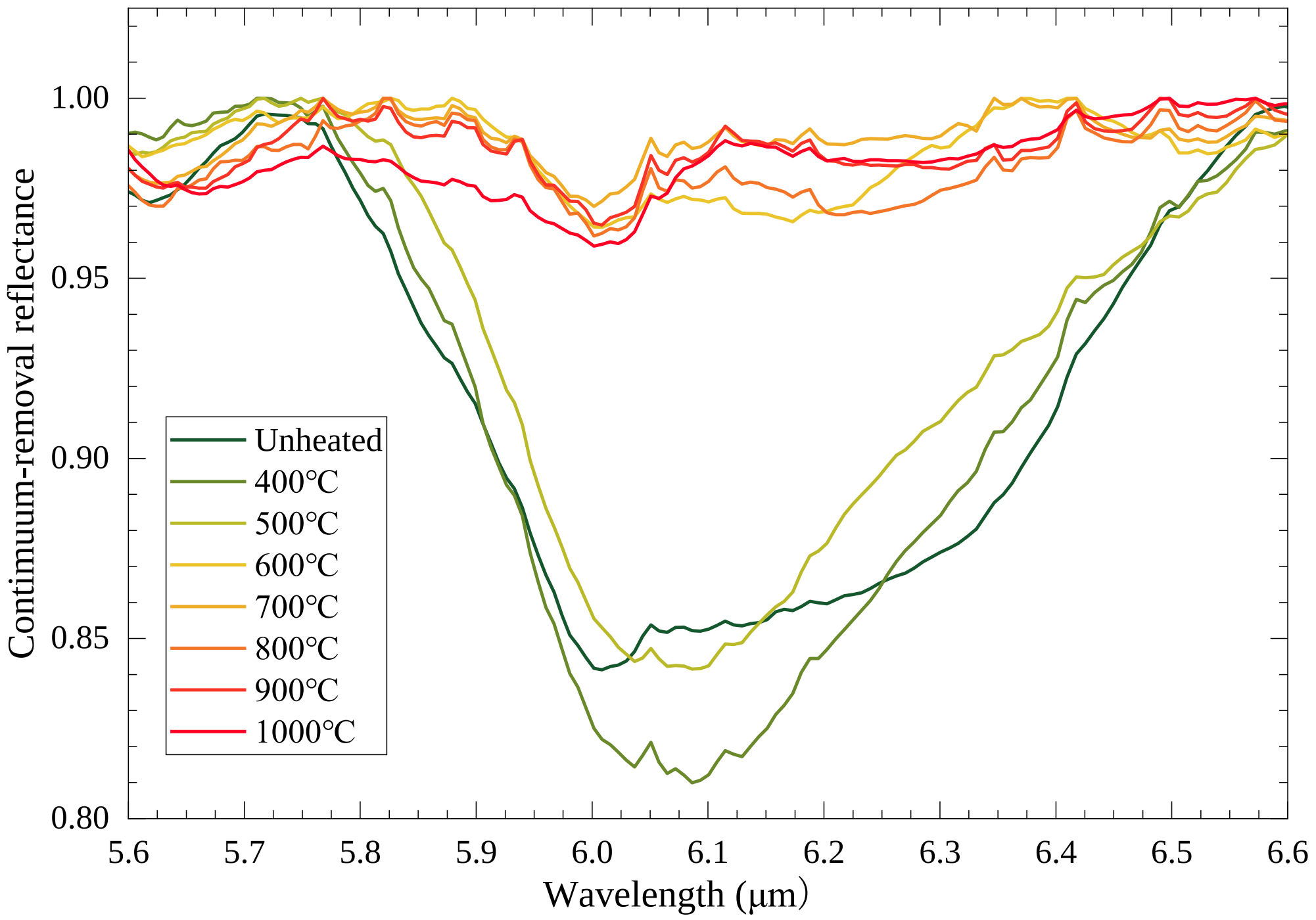

*Figure 4 Continuum-removed spectra of Murchison meteorite heated at different temperatures at 6 μm band.*

During the heating process, the 6 μm absorption band shows no significant variation under 500°C, with an absorption intensity of approximately 15.5% and a band center around 6.07 μm (Figure 3B and Figure 4). After heating above 500°C, however, the absorption band rapidly disappears, leaving a weak absorption feature near 6.01 μm with an intensity of 3.1% only. The different variations observed in the 3 μm (Figure 1B and Figure 2) and 6 μm (Figure 3B and Figure 4) bands after heating can be attributed to the fact that the 6 μm feature is primarily influenced by water molecules, while the 3 μm band is also affected by a variety of minerals. During the heating to 500°C, although the ambient water and some interlayer water in the CCs are removed (Beck et al., 2014), the phyllosilicates undergo only mild metamorphism, and their mineral structure remains relatively intact (Nakamura, 2005; King, 2021). As a result, the 6 μm absorption feature produced by the structural water in the minerals is still present. However, the phyllosilicate minerals undergo dehydration and metamorphism at temperatures above 600°C, leading to recrystallization and the formation of anhydrous silicates (Hiroi et al., 1993; Bates, 2020). After heating to 600–1000°C, a faint absorption becomes apparent at approximately 6.01 μm, which is attributed to terrestrial water in measuring environment or olivine formed from hydrated minerals. These differences allow heated CCs to be distinguished from CCs that are vacuum desiccated or heated only to low temperatures (i.e., <~400°C).

### 3.3. 9–14 μm region

The absorption features produced by the Si-O stretching vibrations in silicates are the most prominent feature in the 9–14 μm region. Different structured silicate minerals exhibit different spectral features and slopes in this region, as well as Reststrahlen bands, transparency features, and Christiansen features (Morlok et al., 2020). These characteristics can be utilized to analyze the proportions of anhydrous silicates and phyllosilicates, along with the degree of aqueous alteration and thermal metamorphism of CCs (Bates et al., 2020b; Morlok et al., 2020).

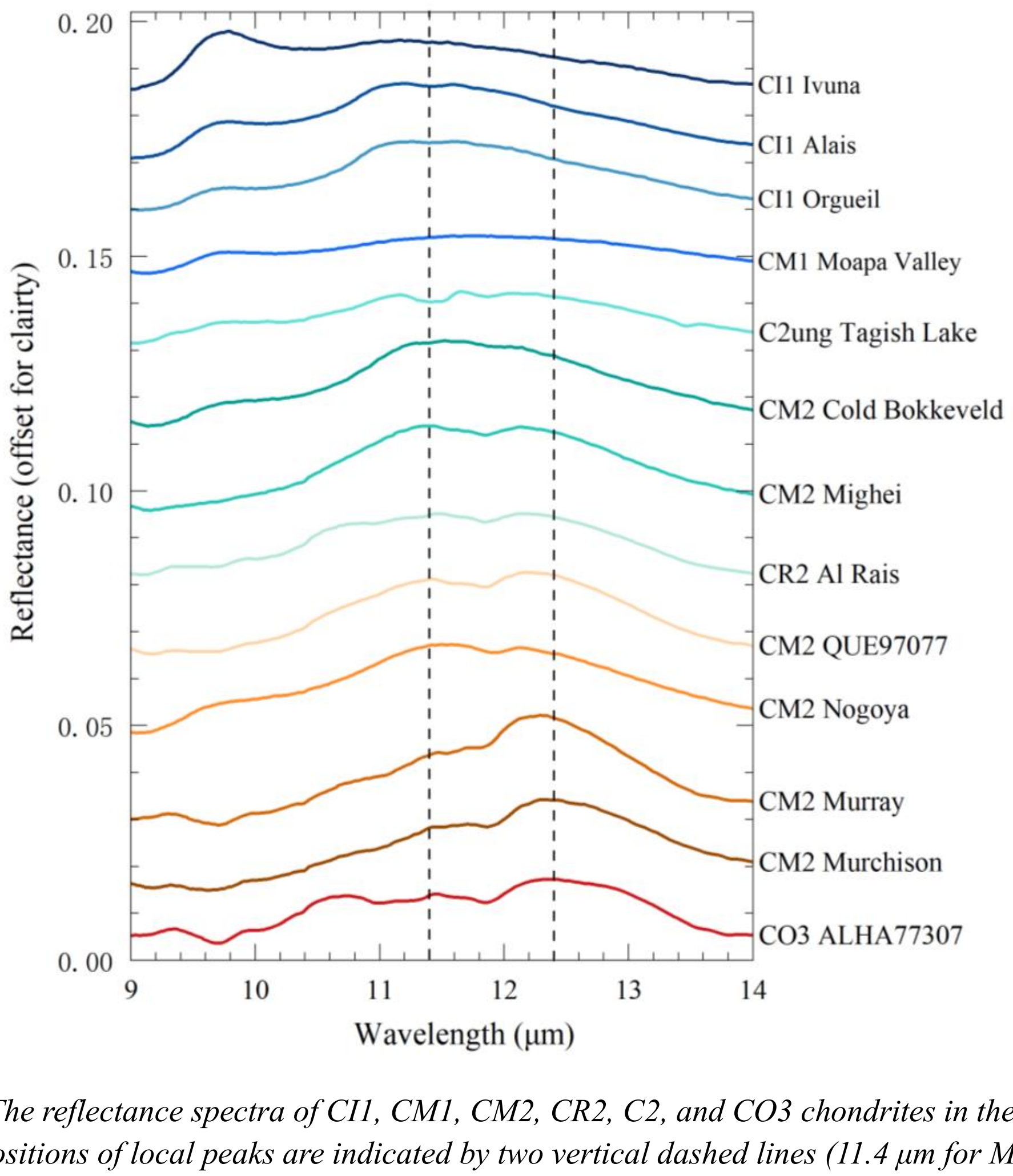

*Figure 5 The reflectance spectra of CI1, CM1, CM2, CR2, C2, and CO3 chondrites in the 9–14 μm. The positions of local peaks are indicated by two vertical dashed lines (11.4 μm for Mg-rich phyllosilicates and 12.4 μm for more olivine present).*

As observed in the spectra of CI1, CM1, CM2, CR2, CO3, and C2ung chondrites in the 9–14 μm range (Figure 5), a shift of the reflectance peak toward shorter wavelengths in this region is associated with an increase in alteration. The center position transitions from around 12.4 μm in low-alterated CM2 Murchison and CO3 ALHA77307 to 11 μm in highly altered CI1 Orgueil and CI1 Alais, and 9.5 μm in CI1 Ivuna. McAdam et al. (2015) report that the Mg-rich phyllosilicates have a shorter wavelength peak near 11.4 μm while olivine presents a single peak at longer wavelengths (~12.3–12.4 μm), and thus the ratio of reflectance at 12.4 μm to 11.4 μm also correlates with the petrological type. As shown in Figure 6, with an increase in the alteration degree the 12.4 μm/11.4 μm ratio decreases for CM2 and CI1 CCs. This is because anhydrous silicates exhibit higher reflectance at 12.4 μm, while phyllosilicates show higher reflectance at 11.4 μm. As the aqueous alteration process progresses, anhydrous silicates in the matrix of CCs gradually transform into phyllosilicates (Beck et al., 2018; McAdam et al., 2015), resulting in a gradual decrease in reflectance at 12.4 μm. Therefore, the reflectance peaks in the 9–14 μm region shift towards shorter wavelengths and the 12.4 μm/11.4 μm ratio decreases with an increase in the alteration degree.

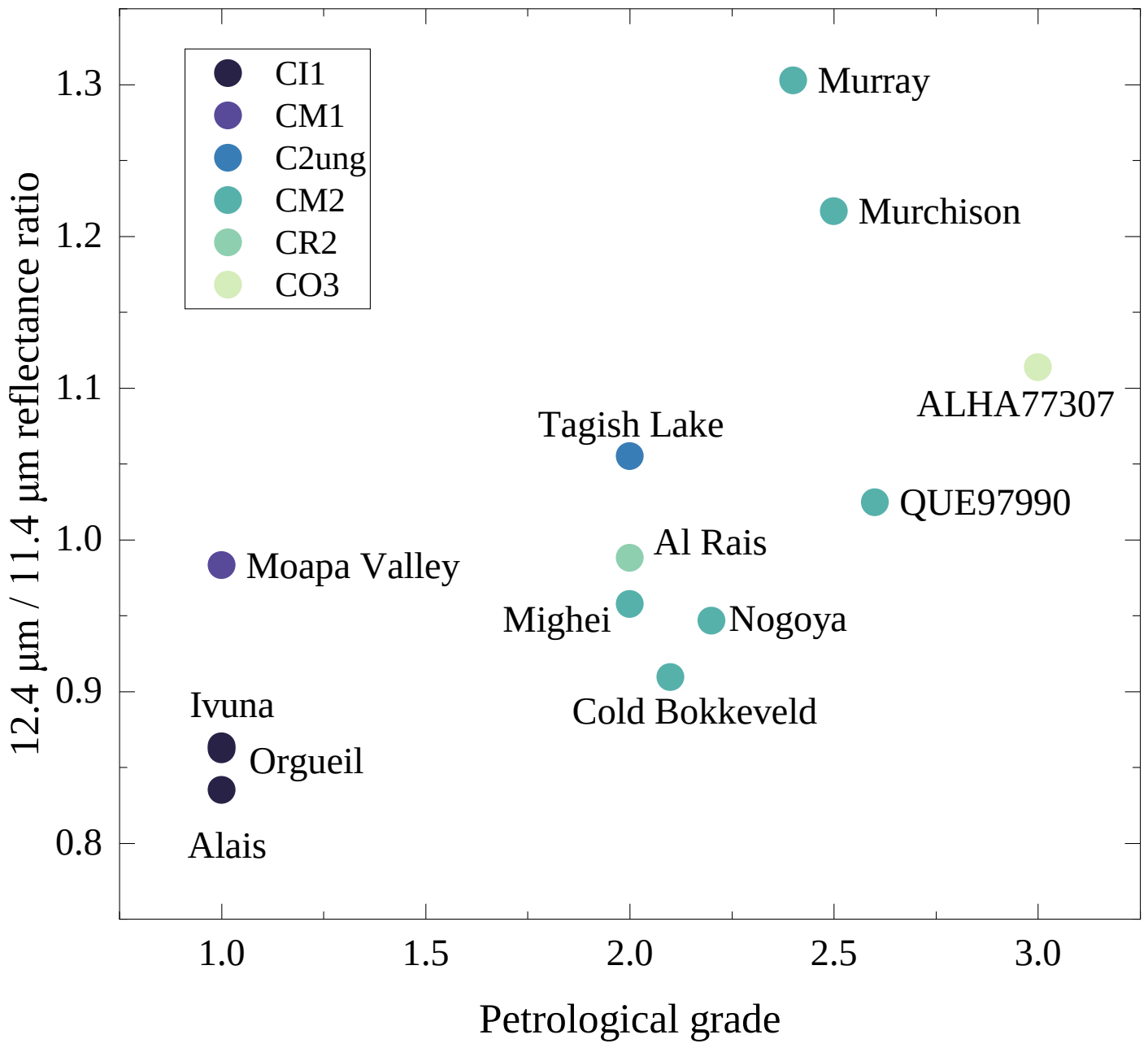

*Figure 6 The relationship between the petrological grade and 12.4 μm/11.4 μm reflectance ratio of CI1, CM1, CM2, CR2, C2, and CO3 chondrites.*

With increasing heating temperature, the reflectance peak in the 9–14 μm range of Murchison meteorite gradually shifts towards longer wavelengths, moving from 10.4 μm to 11.4 μm (Figure 7). The spectral curves of the meteorite heated at 800°C, 900°C, and 1000°C exhibit similarities to CV3 Allende. This phenomenon may be attributed to the progressive recrystallization and metamorphism of phyllosilicates in the CCs, transforming into magnesian olivine. These anhydrous silicates enhance the reflectance in the long-wavelength region, causing the reflectance peak to shift towards longer wavelengths, which has also been reported in emission spectra (Bates et al., 2020a). At temperatures exceeding 800°C, the sample may undergo partial recrystallization to pyroxene and olivine, resulting in more prominent spectral features and a "rougher" curve (more fluctuations, peaks, or irregularities). Notably, CK group samples exhibit distinct spectral features characterized by reflection peaks at approximately 10.9 μm. While most CK meteorites share olivine content with CV and heated CM meteorites, this divergence is attributed to variations in the crystallinity of anhydrous minerals. Unlike the primary olivine found in CK group meteorites, olivine in CM samples originated from poorly crystalline phyllosilicates that transformed into an amorphous phase during heating. This process results in reduced crystallinity and yields a smoother curve which has fewer distinct peaks or fluctuations (Nakamura, 2005; Bates, 2020).

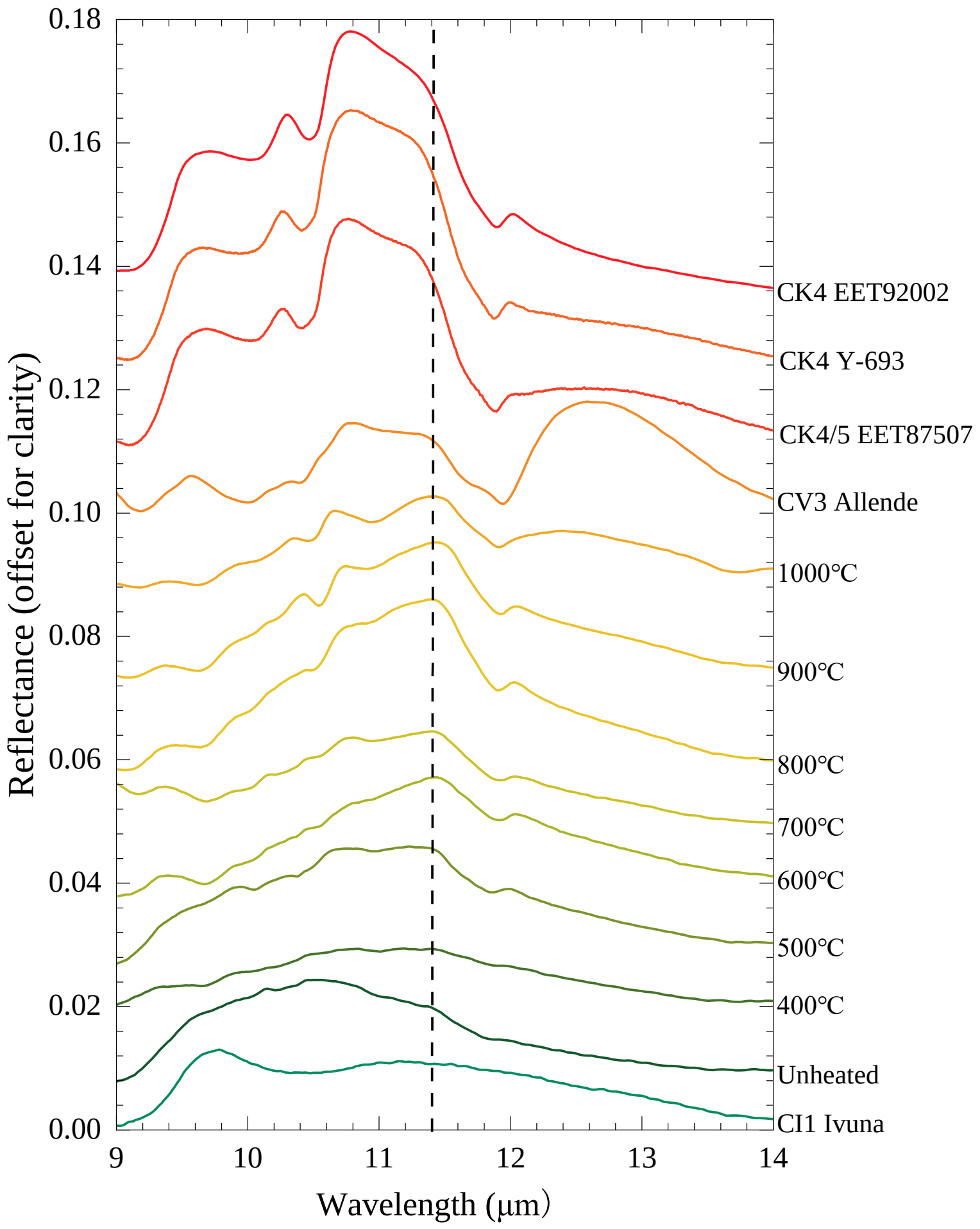

*Figure 7 Comparison of spectral curves at 9–14 μm of CI, CV, and CK group and Murchison samples heated at different temperatures. The dashed line at 11.4 μm is incorporated to highlight the shift in the local reflectance peaks, aiding readers in discerning spectral variation.*

### 3.4. 15–25 μm region

The spectra of heated Murchison samples exhibit strong similarity to CV and CK group chondrites. As shown in Figure 8, the unheated Murchison meteorite displays a reddish slope in 15–25 μm region, with a convex shape and the maximum reflectance around 22.9 μm. After heating at 400°C, the spectral curve becomes flattened. After heating in the range of 500–1000°C, the spectrum gradually exhibits two peaks, with the maximum reflectance observed at 19.5 μm and 24.9 μm, and a concave shape at 22 μm. The spectra of Murchison meteorite heated at 800–1000°C are similar to those of Allende chondrite, while the spectra of CK group chondrites EET92002, Y-693, and EET87507 exhibit similar characteristics with more pronounced dual peaks, with the maximum reflectance located at 18.6 μm and 24.5 μm, respectively. Additionally, the spectra of CK group chondrites show a peak around 16.0 μm. The spectral variations observed in heated Murchison meteorite are attributed to changes in the silicate structure within the meteorite. The presence of phyllosilicate minerals such as antigorite in Murchison meteorite results in a Si-O bending vibration at 21 μm (McAdam et al., 2015). With increasing temperature, the phyllosilicate undergoes progressive dehydration and metamorphism. After heating at 800–1000°C, all phyllosilicates in Murchison meteorite transform into anhydrous silicate minerals such as olivine and pyroxene, exhibiting similar Si-O vibration characteristics (McAdam et al., 2015), resembling the unaltered CV3 Allende. EET92002, Y-693, and EET87507 all belong to the CK group and their parent bodies have generally undergone more extensive thermal metamorphism. As a result, they exhibit higher crystallinity, especially for olivine. This results in a stronger spectral signature in the area.

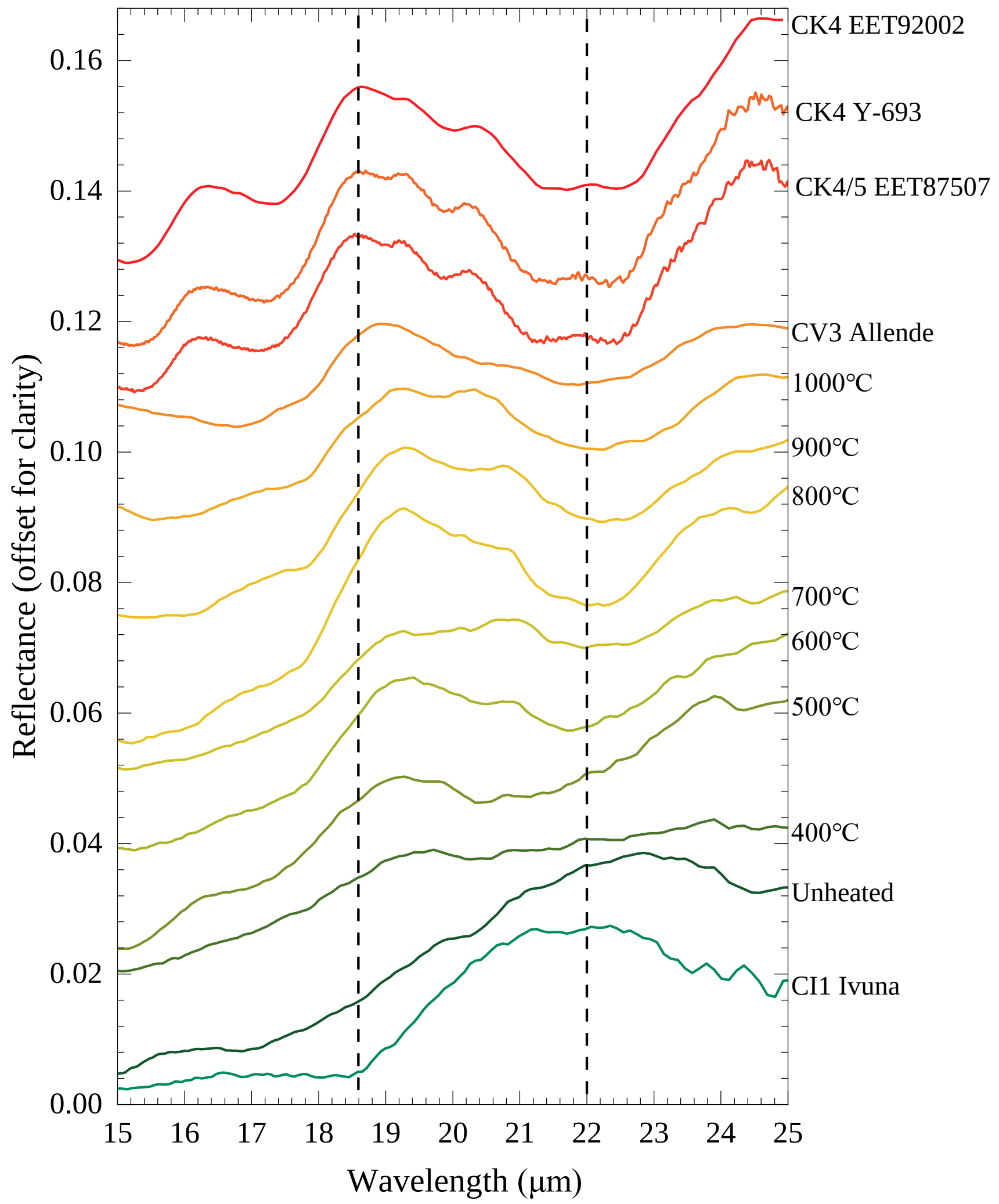

*Figure 8 Comparison of spectral curves at 15–25 μm of CI, CV, and CK group and Murchison samples heated at different temperatures. Dashed lines at 18.6 μm and 22.0 μm are incorporated to highlight the shift in the local reflectance peaks, aiding readers in discerning spectral variation.*

# 4. Link to asteroids

Observations of the 3 μm band in Solar System objects provide the most direct means of studying the aqueous alteration processes on asteroids and the distribution of hydrated materials in the Solar System. However, limitations in the observational capabilities of infrared instruments and the opacity of Earth's atmosphere in the 3 μm region have posed challenges. As a result, the available investigations are mainly focused on large asteroids in the main belt and near-Earth objects (Jones et al., 1990; Bus and Binzel, 2002; Lauretta et al., 2015; Takir and Emery, 2012; Takir et al., 2015; Usui et al., 2019; Takir et al., 2023). Consequently, the available observational data and research on the 3 μm band in asteroids are quite limited.

However, the successful sample return mission, Hayabusa2, has brought surface and subsurface materials from carbonaceous asteroids Ryugu, providing a unique opportunity to analyze 3 μm band spectral data from asteroids. In this section, using the Ryugu as an illustrative example, we will discuss various challenges and limitations encountered when comparing meteorite spectra with asteroid spectra. Additionally, we will analyze the 3 μm band data of mid-belt asteroids and discuss its spectral variation pattern.

### 4.1. Limitations of spectral comparisons between CCs sample and asteroids: terrestrial water, space weathering, and observational effects

As the 3 μm band serves as the primary investigative tool of aqueous alteration, many studies aim to understand asteroids by comparing spectral curve shapes in this region (e.g., Rivkin et al., 2012; Potin et al., 2020; Kitazato et al., 2021; Cantillo et al., 2021; McGraw et al., 2022). However, recent analyses of the Ryugu samples indicate issues with contrasting meteorite spectra with asteroid spectra.

In comparison to CI1 meteorites, spectra of Ryugu and the laboratory samples have lower reflectance values and show a shallower 3 μm-region absorption feature (Kitazato et al., 2019; Pilorget et al., 2022). The 3 μm band center position is consistent with CI1 group meteorites and C2ung Tagish Lake meteorite (2.71–2.72 μm; T. Nakamura et al., 2022), but CI1 group chondrites typically exhibit a 3 μm band depth exceeding 50%. In contrast, band depth of Ryugu samples is around 18% (Pilorget et al., 2022), similar to the 300℃ heated CI1 Orgueil meteorite (Amano et al., 2023), while Ryugu's surface mean spectrum observed by the Hayabusa 2's onboard spectrometer NISR3 is considerably less (~9%; Kitazato et al., 2019), similar to the 800℃ heated CI1 Ivuna meteorite (Kitazato et al., 2021).

This difference between the spectra of the asteroid Ryugu and of the returned samples likely results from the observational effects during the in-situ measurements as well as the space weathering process undergone by the asteroid's surface. On the one hand, there is a substantial impact of grain size, geometrical, and shadow effects of the asteroid's surface on the observed reflectance spectral and absorption features in the 3 μm band (Pilorget et al., 2022; Potin et al., 2022; Yumoto et al., 2023). It might be particularly pronounced on rubble-pile asteroids, where the regolith is dominated by irregular boulders and larger particles, ultimately resulted in a weaker 3 μm band in the Ryugu surface than in the returned finer-grained samples (~9% vs. 18%; Pilorget et al., 2022).

On the other hand, the discrepancy with the CI chondrites' spectra can be attributable to space weathering processes experienced by the original Ryugu material and vacuum desiccation, leading to dehydration at low temperatures; the presence of interlayer water in these samples is notably low (~0.3 wt %; Yokoyama et al., 2022). In contrast, CI chondrites have additional terrestrial water (e.g., adsorbed water, rehydrated water), and secondary hydrated minerals (e.g., hydrous sulfates and ferrihydrites) acquired during terrestrial weathering, which make the 3 μm bands deeper and rounder (Amano et al., 2023; Schultz et al., 2023). The removal of these components through high-temperature heating under reducing conditions aligns the spectra with those of fresh Ryugu samples (Amano et al., 2023).

For the two reasons mentioned above, the observation from Hayabusa 2 has led past studies to erroneously suggest spectral similarities between Ryugu and heated CI1 Ivuna, inferring thermal metamorphism in Ryugu's evolution history (above 300 °C; Kitazato et al., 2021). However, the mineralogical analyses unequivocally indicate similarities with pristine CI chondrites, suggesting Ryugu's thermal history did not exceed 100 °C heating (Ito et al., 2022; E. Nakamura et al., 2022; T. Nakamura et al., 2022; Yokoyama et al., 2022).

In this study, it is essential to acknowledge that all meteorite samples have been influenced by varying amounts of terrestrial water, which significantly affected the spectral features. Previous XRD, IR and water content analyses compared measurements made in ambient conditions with measurements made after pre-heating, revealing the influence of terrestrial water (Beck et al., 2010; King, 2021; Takir et al., 2013); these authors also reported that the samples is immediately contaminated on contact with the atmosphere and rehydrated water is difficult to remove (Matsuoka et al., 2022). While these influences have a significant impact on our ability to precisely measure the degree of alteration and quantitatively analyze material content, it is important to note that the interference of terrestrial water contamination on the spectral features of silicates is relatively limited (Milliken and Mustard, 2007a; Beck et al., 2010); the absorption center position of hydroxyl (-OH) in the 3 μm region depends only on the Fe\Mg content in phyllosilicate, and the MIR regions (9–14 μm and 15-25 μm) are controlled by silicates' vibrational bands. Thus, valuable information about aqueous alteration processes of CCs can still be obtained in the 3 μm region (Hiroi et al., 1996b; Fornasier et al., 1999; Beck et al., 2018). In addition, adsorbed water and pore water in hydrated minerals should not be solely regarded as a terrestrial interference factor. At the present time, sample return missions are limited to near-Earth asteroids that may undergo more pronounced dehydration by space weathering. However, CI-like chondrite parent bodies are expected to contain more water, especially considering that interlayer water should be up to 7 wt % in phyllosilicates formed on Ryugu's parent body (T. Nakamura et al., 2022; Yokoyama et al., 2022). Besides, According to Usui et al. (2019) and Takir et al. (2023), the large asteroids in the outer main belt exhibit rounded 3 μm band features, which may suggest the potential presence of more water content in their surface hydrated materials, and future research endeavors are required to establish the presence and nature of water on these distant bodies.

In summary, changes in band depth and spectral morphology in the 3 μm-region absorption are inevitable due to observational effects, space weathering of asteroids, and contamination from terrestrial water. These factors may potentially lead to erroneous conclusions. Therefore, caution must be exercised when conducting comparative analyses of asteroid and meteorite, or asteroid simulant. It is important to recognize that comparisons of absorption depth and spectral morphology in 3 μm-region and 6 μm-region might not be valid, and quantitative analysis of volatile content in asteroids remains challenging for us. However, relative changes in band positions, especially in the OH absorption region and the mid-infrared region, may hold meaningful information when comparing asteroids and meteorites.

### 4.2. 3 μm band variation in asteroids

In this investigation, our primary focus was to compare the pattern of spectral variations in the 3 μm-region between CCs data (dominated by alteration and metamorphism) and asteroid data acquired through astronomical observations. Here, we do not directly compare the spectral morphology and absorption depths of the 3 μm-region between asteroids and meteorites. Instead, we aim to contribute to a broader goal of unraveling the trends in spectral variations among asteroids and investigating potential explanations for these observed patterns. Therefore, our main objective remains unaffected by the influences mentioned in section 4.1.

The asteroid spectral data were collected from the AKARI observations of 66 asteroids, including 23 C-type asteroids, 17 S-type asteroids, 22 X-type asteroids, 3 D-type asteroids, and 1 V-type asteroid. The asteroid global average spectral data from Hayabusa 2 and OSIRIS-REx were also used for comparison (Hamilton et al., 2019; Kitazato et al., 2019), as well as Ryugu sample data from Amano et al (2023). As revealed by the Ryugu samples, the 3 μm band depth of the surface materials of asteroids in laboratory measurements may be much deeper than from in-space observations. We note that large main-belt asteroids may possess a finer grain size regolith (De Sanctis et al., 2015; Ciarniello et al., 2017; Kurokawa et al., 2020), resulting in smaller deviation from laboratory measurement and astronomical observation. However, we refrain from quantitatively comparing lab measurements of meteorites and astronomical observations of asteroids as a conservative approach. For asteroids without intrinsic hydrous feature, the 3 μm band absorption intensity may be affected by spectral contamination from the stellar background, spectral thermal correction, and continuum removal errors, which are beyond the scope of this paper.

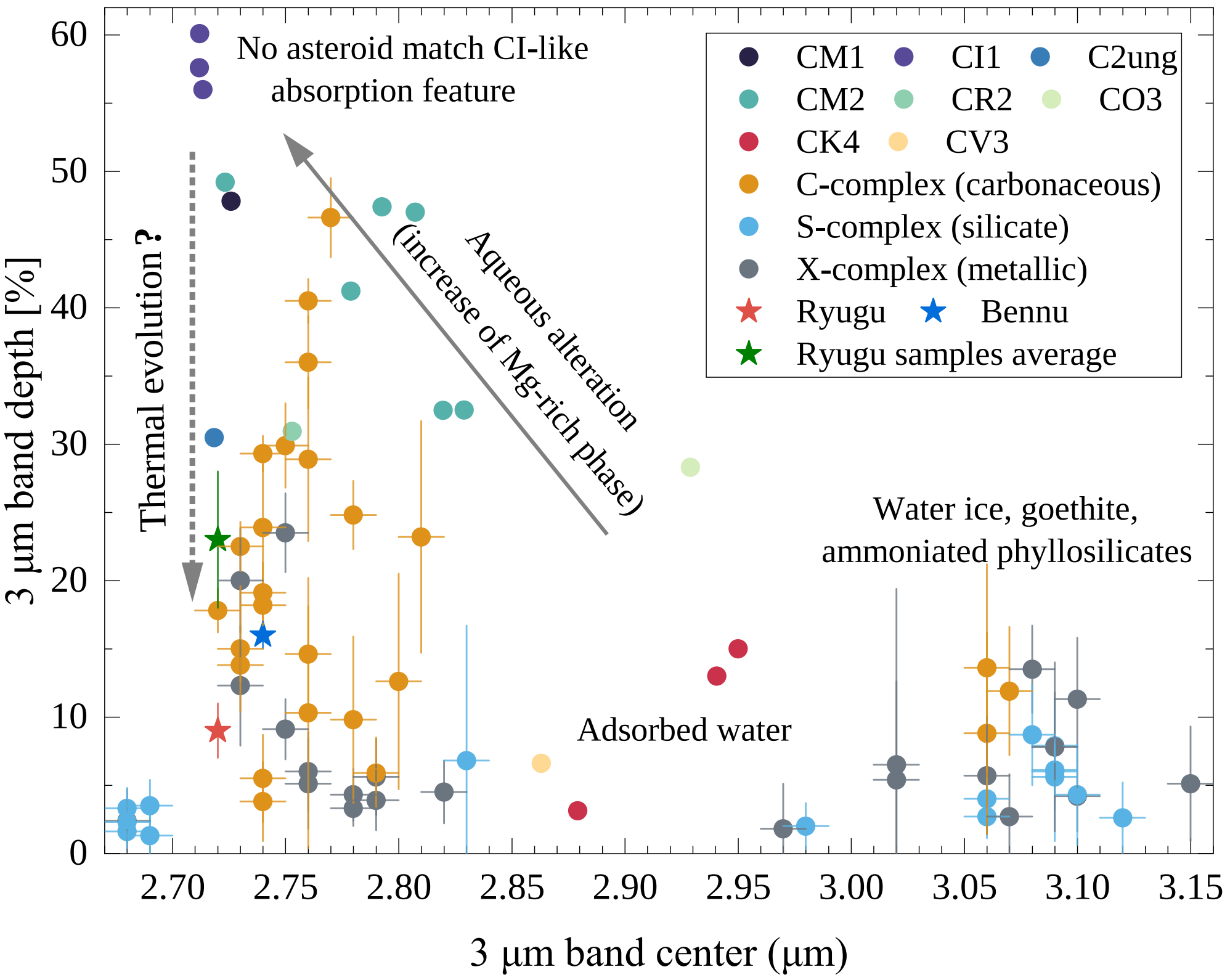

*Figure 9 Comparison of spectral variation trends in the 3 μm bands of CCs and asteroids. Asteroids are grouped based on the SMASSII taxonomy into C-complex (carbonaceous, including C, B, Cb, Cg, Cgh, and D types), S-complex (silicate, including S, A, R, K, L, Sa, Sq, Sr, Sk, Sl, and V types), and X-complex (metal-rich, including X, Xe, Xk, Xc types, corresponding to the M, P, E types in Tholen taxonomy, which lack absorption features and exhibiting high reflectance).*

From Figure 9, it can be observed that the majority of C-complex asteroids have their 3 μm band center ranging from 2.72 μm to 2.85 μm, which is similar to the CM and CI CCs. S-complex asteroids, mostly derived from anhydrous silicates parent bodies, exhibit weak absorption at 2.7 μm (typically <5%) and some of them show absorption below 2.72 μm, which is also attributed to -OH in the hydrated minerals. Some S-complex asteroids show absorption features at 3.1 μm, indicating the possible presence of water ice on their surfaces. This water ice may have originated from other volatile-rich small bodies such as carbonaceous asteroids or comets. While research in Daly et al (2021) has confirmed that solar wind particle bombardment can generate water on S-type asteroids, it remains uncertain whether this process can also give rise to ice on more distant asteroids. X-type asteroids exhibit stronger absorption features at 2.7 μm compared to S-type asteroids, suggesting the presence of hydrated materials. Precise observations of the largest X-type asteroid (Takir et al., 2016), 16 Psyche, have indicated that these hydrated materials may originate from a CC parent body rich in phyllosilicates (Cantillo et al., 2021). All asteroids in our analysis lack the 2.95 μm absorption feature observed in the CK and CO samples. This feature is produced by adsorbed water, and the absence of the 2.95 μm absorption on asteroids suggests that space weathering processes can effectively reduce the content of interlayer water and adsorbed water in rocks, minerals, and particles on asteroid surfaces (Bates et al., 2020a; Noguchi et al., 2023).

The 3 μm absorption features of C-complex asteroids are centered around 2.75 μm, similar to CCs. However, the distribution pattern of the 3 μm band in C-complex asteroids significantly differs from the spectral variations associated with alteration and metamorphism in CCs. Specifically, no asteroid matches CI1 samples' absorption feature, instead, the asteroids with the deepest 3 μm bands have absorption centers near 2.75 μm. Additionally, there appears to be a subtle trend indicating that as the band center position shifts towards shorter wavelengths, the absorption depth decreases from 45% to 13%. Usui et al. (2019) first reported this trend and suggested an explanation for this trend during their analysis of AKARI data, suggesting that this trend is caused by the dehydration process in the CCs parent bodies. The phyllosilicates decrease due to thermal metamorphism, and interlayer cations are gradually replaced by magnesium, resulting in the shift of the band center to shorter wavelengths and a decrease in band depth.

Although this trend can be influenced by the observational biases and asteroid size limitations, it is worth noting that in the heating experiment on Murchison  and other studies involving heating experiment using Mg-rich phyllosilicate minerals and other materials, such as CI1 chondrites, brucite, and serpentine showed that the absorption peak generated by -OH at 2.72 μm does not shift during the heating process (Hiroi et al., 1996a; Milliken and Mustard, 2007b; Kitazato et al., 2019; Hiroi et al., 2021; Matsuoka et al., 2022). Therefore, the results of our study suggest that the Usui et al. (2019) explanation may not be accurate. The differences observed in the AKARI data could be attributed to variations in the early formation and evolution processes of CCs parent asteroids. Highly altered chondrites, such as type 1 CCs, often originate from specific types of planetesimals characterized by similar water-to-rock ratios, radius, and short-lived isotope compositions. These planetesimals undergo extensive aqueous alteration, transforming all original silicates into phyllosilicates. In contrast, the parent asteroids of CM2 group CCs experienced subsequent prolonged heating processes, typically undergoing more pronounced thermal metamorphism after the aqueous alteration phase during the planetesimal stage (Gail et al., 2014; Weidenschilling, 2019). On the other hand, CI1 and CM1 chondrites collected on Earth are remnants of these planetesimals after early collisional processes, which have not undergone metamorphic effects.

### 4.3. Applications for upcoming exploration

Vernazza et al. (2012) proposed the use of spinel feature near 14–15 μm to indicate thermal metamorphism, but also noted that these feature does not seem to be evident in the RELAB samples. Therefore, simulated asteroid materials based on carbonaceous chondrites may offer further constraints about the nature of asteroid surface composition and surface structure (Britt et al., 2019), such as developing quantitative indices to assess the precise relationship between mineralogical characterization and spectral features in the 15–25 μm and longer wavelength range (Fabian et al., 2001), and understanding the effect of porosity and grain size on MIR spectral behavior (Morlok et al., 2020; Vernazza et al., 2012).

In the next decade, it is expected that space infrared telescopes such as the NASA James Webb Space Telescope (JWST) and main belt exploration missions will acquire more near-infrared and mid-infrared spectral data of C-complex asteroids. Comparing these data with the infrared spectral variations of CCs in the 3 μm, 9–14 μm, and 15–25 μm regions provides valuable insights into the volatile component content and thermal evolution history of asteroids. In addition, the MIR region can visualize the aqueous alteration and thermal metamorphism trends of asteroids, and this region is not subject to severe atmospheric interference, so that ground-based infrared telescope observations of asteroids should be attempted in the future.

## 5. Conclusions

The reflectance spectra of carbonaceous chondrites are complex, and the spectral features of aqueous alteration are influenced by several factors. The most prominent features related to aqueous alteration are the 3 μm band and the 6 μm band. The spectral shapes in the 9–14 μm and 15–25 μm regions can reflect the proportion of phyllosilicate minerals in the CCs. This study investigates the spectral variations during aqueous alteration and thermal metamorphism of CCs through spectral analysis of 17 samples with different petrological types and 1 CC heating experiment.

The 3 μm band center and depth are correlated, indicating an increase in hydrous alteration minerals (e.g., magnesium-rich phyllosilicate) as the degree of alteration increases. During thermal metamorphism, the depth of the 3 μm band decreases and the absorption center shifts to longer wavelengths, indicating the dehydration and recrystallization of hydrated minerals.

The 6 μm band depth increases with aqueous alteration, reflecting an overall increase in the total water content of CCs during the alteration process. Compared to the 3 μm band, the center position shows a smaller variation for different CCs and less correlated to the aqueous alteration. During heating process, the 6 μm band disappears at temperatures of 600°C and above, indicating the dehydration of interlayer water and mineral-bound water in the CCs.

The 9–14 μm region reflect the alteration degree and the content of phyllosilicates in CCs. As the alteration increases, the content of phyllosilicates increases, resulting in a decrease in the reflectance ratio of 12.4 μm/11.4 μm and a shift of the reflectance peak towards shorter wavelengths. During thermal metamorphism, the anhydrous silicate content increases, causing the reflectance peak to shift towards longer wavelengths with increasing temperature. However, the spectral feature in this region for CK samples is significantly different from other CCs due to the higher crystallinity of the anhydrous minerals.

The 15–25 μm region indicates the degree of metamorphism in hydrated CCs. With increasing heating temperature, the reflection peak around 22 μm gradually transforms into a double peak located at 18.6 μm and 24.5 μm, like CV chondrites. This transformation may be attributed to the formation of olivine through thermal metamorphism. The double peak feature is more pronounced in CK chondrites, possibly due to more intense thermal metamorphism experienced by their parent bodies.

When comparing the 3 μm band variation pattern of CCs to main-belt asteroids, the general lack of asteroids matched to highly altered meteorites may indicate that CI-like carbonaceous chondrites parent bodies are rare or underwent more intense thermal metamorphism.

# Acknowledgment

We are very grateful to Dr. Ashley King and Dr. Helena Bates for their help in interpreting the spectral features and for supporting this study, and thank two anonymous reviewers for constructive comments. This work was supported by the Strategic Priority Research Program of Chinese Academy of Sciences (XDB41 000000), the National Natural Sciences Foundation of China (Grant Nos. 12150009, 62227901, 11633009), the Pre-research project on Civil Aerospace Technologies by CNSA (Grant Nos. KJSP2020020204, KJSP2020020205, and KJSP2020020102), the Minor Planet Foundation, JSPS KAKENHI (Grant Nos. 20KK0080, 21H04514, 21K13976, 22H01290, 22H05150), the Space debris and NEO research project (Grant Nos. KJSP2020020204, KJSP2020020205), and Canadian Space Agency (Grant No. 22EXPOSIWI), and the Canada Foundation for Innovation and Research Manitoba (Grant Nos. CFI 1504 and 2450) and the University of Winnipeg. We would like to acknowledge the NASA RELAB database as a vital resource to the planetary science community.

# Reference

Alexander, C.M.O., Bowden, R., Fogel, M.L., Howard, K.T., Herd, C.D.K., Nittler, L.R., 2012. The Provenances of Asteroids, and Their Contributions to the Volatile Inventories of the Terrestrial Planets. Science 337, 721–723. https://doi.org/10.1126/science.1223474

Amano, K., Matsuoka, M., Nakamura, T., Kagawa, E., Fujioka, Y., Potin, S.M., Hiroi, T., Tatsumi, E., Milliken, R.E., Quirico, E., Beck, P., Brunetto, R., Uesugi, M., Takahashi, Y., Kawai, T., Yamashita, S., Enokido, Y., Wada, T., Furukawa, Y., Zolensky, M.E., Takir, D., Domingue, D.L., Jaramillo-Correa, C., Vilas, F., Hendrix, A.R., Kikuiri, M., Morita, T., Yurimoto, H., Noguchi, T., Okazaki, R., Yabuta, H., Naraoka, H., Sakamoto, K., Tachibana, S., Yada, T., Nishimura, M., Nakato, A., Miyazaki, A., Yogata, K., Abe, M., Okada, T., Usui, T., Yoshikawa, M., Saiki, T., Tanaka, S., Terui, F., Nakazawa, S., Watanabe, S., Tsuda, Y., 2023. Reassigning CI chondrite parent bodies based on reflectance spectroscopy of samples from carbonaceous asteroid Ryugu and meteorites. Science Advances 9, eadi3789. https://doi.org/10.1126/sciadv.adi3789

Bates, H., 2020. Understanding the formation and evolution of asteroids through analysis of carbonaceous chondrites (http://purl.org/dc/dcmitype/Text). University of Oxford.

Bates, H.C., Hanna, K.L.D., King, A.J., Bowles, N.E., Russell, S.S., 2020a. A Spectral Investigation of Aqueously and Thermally Altered CM, CM-An, and CY Chondrites Under Simulated Asteroid Conditions for Comparison With OSIRIS-REx and Hayabusa2 Observations. Journal of Geophysical Research 30.

Bates, H.C., King, A.J., Hanna, K.L.D., Bowles, N.E., Russell, S.S., 2020b. Linking mineralogy and spectroscopy of highly aqueously altered CM and CI carbonaceous chondrites in preparation for primitive asteroid sample return. Meteoritics & Planetary Science 55, 77–101.

Beck, P., Garenne, A., Quirico, E., Bonal, L., Montes-Hernandez, G., Moynier, F., Schmitt, B., 2014. Transmission infrared spectra (2–25μm) of carbonaceous chondrites (CI, CM, CV–CK, CR, C2 ungrouped): Mineralogy, water, and asteroidal processes. Icarus 229, 263–277. https://doi.org/10.1016/j.icarus.2013.10.019

Beck, P., Maturilli, A., Garenne, A., Vernazza, P., Helbert, J., Quirico, E., Schmitt, B., 2018. What is controlling the reflectance spectra (0.35–150 μm) of hydrated (and dehydrated) carbonaceous chondrites? Icarus 313, 124–138. https://doi.org/10.1016/j.icarus.2018.05.010

Beck, P., Quirico, E., Montes-Hernandez, G., Bonal, L., Bollard, J., Orthous-Daunay, F.-R., Howard, K.T., Schmitt, B., Brissaud, O., Deschamps, F., Wunder, B., Guillot, S., 2010. Hydrous mineralogy of CM and CI chondrites from infrared spectroscopy and their relationship with low albedo asteroids. Geochimica et Cosmochimica Acta 74, 4881–4892. https://doi.org/10.1016/j.gca.2010.05.020

Beck, P., Quirico, E., Sevestre, D., Montes-Hernandez, G., Pommerol, A., Schmitt, B., 2011. Goethite as an alternative origin of the 3.1 $\mu$ m band on dark asteroids. A&A 526, A85. https://doi.org/10.1051/0004-6361/201015851

Binzel, R.P., Gehrels, T., Matthews, M.S., 1989. Asteroids II, Asteroids II. Tucso AZ, University of Arizona Press.

Bishop, J.L., Lane, M.D., Dyar, M.D., Brown, A.J., 2008. Reflectance and emission spectroscopy study of four groups of phyllosilicates: smectites, kaolinite-serpentines, chlorites and micas. Clay Minerals 43, 35–54. https://doi.org/10.1180/claymin.2008.043.1.03

Bottke, W.F., DeMeo, F.E., Michel, P., 2015. Asteroids IV. University of Arizona Press, Tucson.

Braukmüller, N., Wombacher, F., Hezel, D.C., Escoube, R., Münker, C., 2018. The chemical composition of carbonaceous chondrites: Implications for volatile element depletion, complementarity and alteration. Geochimica et Cosmochimica Acta 239, 17–48. https://doi.org/10.1016/j.gca.2018.07.023

Brearley, A.J., 1993. Matrix and fine-grained rims in the unequilibrated CO3 chondrite, ALHA77307: Origins and evidence for diverse, primitive nebular dust components. Geochimica et Cosmochimica Acta 57, 1521–1550. https://doi.org/10.1016/0016-7037(93)90011-K

Britt, D.T., Cannon, K.M., Donaldson Hanna, K., Hogancamp, J., Poch, O., Beck, P., Martin, D., Escrig, J., Bonal, L., Metzger, P.T., 2019. Simulated asteroid materials based on carbonaceous chondrite mineralogies. Meteoritics & Planetary Science 54, 2067–2082. https://doi.org/10.1111/maps.13345

Browning, L.B., McSween, H.Y., Zolensky, M.E., 1996. Correlated alteration effects in CM carbonaceous chondrites. Geochimica et Cosmochimica Acta 60, 2621–2633. https://doi.org/10.1016/0016-7037(96)00121-4

Bühler, R.W., 1988. Meteorite. Birkhäuser, Basel. https://doi.org/10.1007/978-3-0348-6667-5

Bus, S.J., Binzel, R.P., 2002. Phase II of the Small Main-Belt Asteroid Spectroscopic Survey. A Feature-Based Taxonomy. Icarus 158, 146–177. https://doi.org/10.1006/icar.2002.6856

Campins, H., Hargrove, K., Pinilla-Alonso, N., Howell, E.S., Kelley, M.S., Licandro, J., Mothé-Diniz, T., Fernández, Y., Ziffer, J., 2010. Water ice and organics on the surface of the asteroid 24 Themis. Nature 464, 1320–1321. https://doi.org/10.1038/nature09029

Cantillo, D.C., Reddy, V., Sharkey, B.N.L., Pearson, N.A., Sanchez, J.A., Izawa, M.R.M., Kareta, T., Campbell, T.S., Chabra, O., 2021. Constraining the Regolith Composition of Asteroid (16) Psyche via Laboratory Visible Near-infrared Spectroscopy. Planet. Sci. J. 2, 95. https://doi.org/10.3847/PSJ/abf63b

Ciarniello, M., Sanctis, M.C.D., Ammannito, E., Raponi, A., Longobardo, A., Palomba, E., Carrozzo, F.G., Tosi, F., Li, J.-Y., Schröder, S.E., Zambon, F., Frigeri, A., Fonte, S., Giardino, M., Pieters, C.M., Raymond, C.A., Russell, C.T., 2017. Spectrophotometric properties of dwarf planet Ceres from the VIR spectrometer on board the Dawn mission. A&A 598, A130. https://doi.org/10.1051/0004-6361/201629490

Daly, L., Lee, M.R., Hallis, L.J., Ishii, H.A., Bradley, J.P., Bland, P.A., Saxey, D.W., Fougerouse, D., Rickard, W.D.A., Forman, L.V., Timms, N.E., Jourdan, F., Reddy, S.M., Salge, T., Quadir, Z., Christou, E., Cox, M.A., Aguiar, J.A., Hattar, K., Monterrosa, A., Keller, L.P., Christoffersen, R., Dukes, C.A., Loeffler, M.J., Thompson, M.S., 2021. Solar wind contributions to Earth's oceans. Nat Astron 5, 1275–1285. https://doi.org/10.1038/s41550-021-01487-w

De Sanctis, M.C., Ammannito, E., Raponi, A., Marchi, S., McCord, T.B., McSween, H.Y., Capaccioni, F., Capria, M.T., Carrozzo, F.G., Ciarniello, M., Longobardo, A., Tosi, F., Fonte, S., Formisano, M., Frigeri, A., Giardino, M., Magni, G., Palomba, E., Turrini, D., Zambon, F., Combe, J.-P., Feldman, W., Jaumann, R., McFadden, L.A., Pieters, C.M., Prettyman, T., Toplis, M., Raymond, C.A., Russell, C.T., 2015. Ammoniated phyllosilicates with a likely outer Solar System origin on (1) Ceres. Nature 528, 241–244. https://doi.org/10.1038/nature16172

Donaldson Hanna, K.L., Schrader, D.L., Cloutis, E.A., Cody, G.D., King, A.J., McCoy, T.J., Applin, D.M., Mann, J.P., Bowles, N.E., Brucato, J.R., Connolly, H.C., Dotto, E., Keller, L.P., Lim, L.F., Clark, B.E., Hamilton, V.E., Lantz, C., Lauretta, D.S., Russell, S.S., Schofield, P.F., 2019. Spectral characterization of analog samples in anticipation of OSIRIS-REx's arrival at Bennu: A blind test study. Icarus 319, 701–723. https://doi.org/10.1016/j.icarus.2018.10.018

Duan, A., Wu, Y., Cloutis, E.A., Yu, J., Li, S., Jiang, Y., 2021. Heating of carbonaceous materials: Insights into the effects of thermal metamorphism on spectral properties of carbonaceous chondrites and asteroids. Meteoritics & Planetary Science 56, 2035–2046. https://doi.org/10.1111/maps.13750

Fabian, D., Posch, T., Mutschke, H., Kerschbaum, F., Dorschner, J., 2001. Infrared optical properties of spinels - A study of the carrier of the 13, 17 and 32 µm emission features observed in ISO-SWS spectra of oxygen-rich AGB stars. A&A 373, 1125–1138. https://doi.org/10.1051/0004-6361:20010657

Fumihiko, U., Hasegawa, S., Takafumi, O., Takashi, O., 2019. AKARI/IRC near-infrared asteroid spectroscopic survey: AcuA-spec. Publications of the Astronomical Society of Japan Volume 71. https://doi.org/10.1093/pasj/psy125

Gail, H., Trieloff, M., Breuer, D., Spohn, T., 2014. Early Thermal Evolution of Planetesimals and Its Impact on Processing and Dating of Meteoritic Material. pp. 571–593. https://doi.org/10.2458/azu_uapress_9780816531240-ch025

Galiano, A., Palomba, E., D'Amore, M., Zinzi, A., Dirri, F., Longobardo, A., Kitazato, K., Iwata, T., Matsuoka, M., Hiroi, T., Takir, D., Nakamura, T., Abe, M., Ohtake, M., Matsuura, S., Watanabe, S., Yoshikawa, M., Saiki, T., Tanaka, S., Okada, T., Yamamoto, Y., Takei, Y., Shirai, K., Hirata, N., Hirata, N., Matsumoto, K., Tsuda, Y., 2020. Characterization of the Ryugu surface by means of the variability of the near-infrared spectral slope in NIRS3 data. Icarus 351, 113959. https://doi.org/10.1016/j.icarus.2020.113959

Garenne, A., Beck, P., Montes-Hernandez, G., Chiriac, R., Toche, F., Quirico, E., Bonal, L., Schmitt, B., 2014. The abundance and stability of "water" in type 1 and 2 carbonaceous chondrites (CI, CM and CR). Geochimica et Cosmochimica Acta 20.

Hamilton, V.E., Simon, A.A., Christensen, P.R., Reuter, D.C., Clark, B.E., Barucci, M.A., Bowles, N.E., Boynton, W.V., Brucato, J.R., Cloutis, E.A., Connolly, H.C., Donaldson Hanna, K.L., Emery, J.P., Enos, H.L., Fornasier, S., Haberle, C.W., Hanna, R.D., Howell, E.S., Kaplan, H.H., Keller, L.P., Lantz, C., Li, J.-Y., Lim, L.F., McCoy, T.J., Merlin, F., Nolan, M.C., Praet, A., Rozitis, B., Sandford, S.A., Schrader, D.L., Thomas, C.A., Zou, X.-D., Lauretta, D.S., 2019. Evidence for widespread hydrated minerals on asteroid (101955) Bennu. Nat Astron 3, 332–340. https://doi.org/10.1038/s41550-019-0722-2

Hiroi, T., Kaiden, H., Imae, N., Misawa, K., Kojima, H., Sasaki, S., Matsuoka, M., Nakamura, T., Bish, D.L., Ohtsuka, K., Howard, K.T., Robertson, K.R., Milliken, R.E., 2021. UV-visible-infrared spectral survey of Antarctic carbonaceous chondrite chips. Polar Science, Special Issue on "The Sixth International Symposium on Arctic Research (ISAR-6)" 29, 100723. https://doi.org/10.1016/j.polar.2021.100723

Hiroi, T., Pieters, C.M., Zolensky, M.E., Lipschutz, M.E., 1993. Evidence of Thermal Metamorphism on the C, G, B, and F Asteroids. Science 261, 1016–1018. https://doi.org/10.1126/science.261.5124.1016

Hiroi, T., Pieters, C.M., Zolensky, M.E., Prinz, M., 1996a. Reflectance Spectra (UV-3 micrometers) of Heated Ivuna (CI) Meteorite and Newly Identified Thermally Metamorphosed CM Chondrites 27, 551.

Hiroi, T., Zolensky, M.E., Pieters, C.M., Lipschutz, M.E., 1996b. Thermal metamorphism of the C, G, B, and F asteroids seen from the 0.7 micron, 3 micron and UV absorption strengths in comparison with carbonaceous chondrites. Meteoritics and Planetary Science 31, 321–327. https://doi.org/10.1111/j.1945-5100.1996.tb02068.x

Honniball, C.I., 2021. Molecular water detected on the sunlit Moon by SOFIA. Nature Astronomy 5, 13.

Howard, K.T., Alexander, C.M.O., Schrader, D.L., Dyl, K.A., 2015. Classification of hydrous meteorites (CR, CM and C2 ungrouped) by phyllosilicate fraction: PSD-XRD modal mineralogy and planetesimal environments. Geochimica et Cosmochimica Acta 149, 206–222. https://doi.org/10.1016/j.gca.2014.10.025

Howard, K.T., Benedix, G.K., Bland, P.A., Cressey, G., 2011. Modal mineralogy of CM chondrites by X-ray diffraction (PSD-XRD): Part 2. Degree, nature and settings of aqueous alteration. Geochimica et Cosmochimica Acta 75, 2735–2751. https://doi.org/10.1016/j.gca.2011.02.021

Howard, K.T., Benedix, G.K., Bland, P.A., Cressey, G., 2009. Modal mineralogy of CM2 chondrites by X-ray diffraction (PSD-XRD). Part 1: Total phyllosilicate abundance and the degree of aqueous alteration. Geochimica et Cosmochimica Acta 73, 4576–4589. https://doi.org/10.1016/j.gca.2009.04.038

Huss, G.R., Rubin, A.E., Grossman, J.N., 2006. Thermal Metamorphism in Chondrites, Meteorites and the Early Solar System II.

Ito, M., Tomioka, N., Uesugi, M., Yamaguchi, A., Shirai, N., Ohigashi, T., Liu, M.-C., Greenwood, R.C., Kimura, M., Imae, N., Uesugi, K., Nakato, A., Yogata, K., Yuzawa, H., Kodama, Y., Tsuchiyama, A., Yasutake, M., Findlay, R., Franchi, I.A., Malley, J.A., McCain, K.A., Matsuda, N., McKeegan, K.D., Hirahara, K., Takeuchi, A., Sekimoto, S., Sakurai, I., Okada, I., Karouji, Y., Arakawa, M., Fujii, A., Fujimoto, M., Hayakawa, M., Hirata, Naoyuki, Hirata, Naru, Honda, R., Honda, C., Hosoda, S., Iijima, Y., Ikeda, H., Ishiguro, M., Ishihara, Y., Iwata, T., Kawahara, K., Kikuchi, S., Kitazato, K., Matsumoto, K., Matsuoka, M., Michikami, T., Mimasu, Y., Miura, A., Mori, O., Morota, T., Nakazawa, S., Namiki, N., Noda, H., Noguchi, R., Ogawa, N., Ogawa, K., Okada, T., Okamoto, C., Ono, G., Ozaki, M., Saiki, T., Sakatani, N., Sawada, H., Senshu, H., Shimaki, Y., Shirai, K., Sugita, S., Takei, Y., Takeuchi, H., Tanaka, S., Tatsumi, E., Terui, F., Tsukizaki, R., Wada, K., Yamada, M., Yamada, T., Yamamoto, Y., Yano, H., Yokota, Y., Yoshihara, K., Yoshikawa, M., Yoshikawa, K., Fukai, R., Furuya, S., Hatakeda, K., Hayashi, T., Hitomi, Y., Kumagai, K., Miyazaki, A., Nishimura, M., Soejima, H., Iwamae, A., Yamamoto, D., Yoshitake, M., Yada, T., Abe, M., Usui, T., Watanabe, S., Tsuda, Y., 2022. A pristine record of outer Solar System materials from asteroid Ryugu’s returned sample. Nat Astron 6, 1163–1171. https://doi.org/10.1038/s41550-022-01745-5

Jones, T.D., Lebofsky, L.A., Lewis, J.S., Marley, M.S., 1990. The composition and origin of the C, P, and D asteroids: Water as a tracer of thermal evolution in the outer belt. Icarus 88, 172–192. https://doi.org/10.1016/0019-1035(90)90184-B

Kameda, S., Yokota, Y., Kouyama, T., Tatsumi, E., Ishida, M., Morota, T., Honda, R., Sakatani, N., Yamada, M., Matsuoka, M., Suzuki, H., Cho, Y., Hayakawa, M., Honda, C., Sawada, H., Yoshioka, K., Ogawa, K., Sugita, S., 2021. Improved method of hydrous mineral detection by latitudinal distribution of 0.7-μm surface reflectance absorption on the asteroid Ryugu. Icarus 360, 114348. https://doi.org/10.1016/j.icarus.2021.114348

Kaplan, H.H., Milliken, R.E., Alexander, C.M.O., 2018. New Constraints on the Abundance and Composition of Organic Matter on Ceres. Geophysical Research Letters 45, 5274–5282. https://doi.org/10.1029/2018GL077913

Kaplan, H.H., Milliken, R.E., Alexander, C.M.O., Herd, C.D.K., 2019. Reflectance spectroscopy of insoluble organic matter (IOM) and carbonaceous meteorites. Meteoritics & Planetary Science 54, 1051–1068. https://doi.org/10.1111/maps.13264

King, A.J., 2021. Thermal alteration of CM carbonaceous chondrites: Mineralogical changes and metamorphic temperatures. Geochimica et Cosmochimica Acta 24.

King, A.J., Schofield, P.F., Howard, K.T., Russell, S.S., 2015. Modal mineralogy of CI and CI-like chondrites by X-ray diffraction. Geochimica et Cosmochimica Acta 165, 148–160. https://doi.org/10.1016/j.gca.2015.05.038

King, A.J., Schofield, P.F., Russell, S.S., 2017. Type 1 aqueous alteration in CM carbonaceous chondrites: Implications for the evolution of water-rich asteroids. Meteoritics & Planetary Science 52, 1197–1215. https://doi.org/10.1111/maps.12872

Kitazato, K., Milliken, R.E., Iwata, T., Abe, M., Ohtake, M., Matsuura, S., Arai, T., Nakauchi, Y., Nakamura, T., Matsuoka, M., Senshu, H., Hirata, N., Hiroi, T., Pilorget, C., Brunetto, R., Poulet, F., Riu, L., Bibring, J.-P., Takir, D., Domingue, D.L., Vilas, F., Barucci, M.A., Perna, D., Palomba, E., Galiano, A., Tsumura, K., Osawa, T., Komatsu, M., Nakato, A., Arai, T., Takato, N., Matsunaga, T., Takagi, Y., Matsumoto, K., Kouyama, T., Yokota, Y., Tatsumi, E., Sakatani, N., Yamamoto, Y., Okada, T., Sugita, S., Honda, R., Morota, T., Kameda, S., Sawada, H., Honda, C., Yamada, M., Suzuki, H., Yoshioka, K., Hayakawa, M., Ogawa, K., Cho, Y., Shirai, K., Shimaki, Y., Hirata, N., Yamaguchi, A., Ogawa, N., Terui, F., Yamaguchi, T., Takei, Y., Saiki, T., Nakazawa, S., Tanaka, S., Yoshikawa, M., Watanabe, S., Tsuda, Y., 2019. The surface composition of asteroid 162173 Ryugu from Hayabusa2 near-infrared spectroscopy. Science 364, 272–275. https://doi.org/10.1126/science.aav7432

Kitazato, K., Milliken, R.E., Iwata, T., Abe, M., Ohtake, M., Matsuura, S., Takagi, Y., Nakamura, T., Hiroi, T., Matsuoka, M., Riu, L., Nakauchi, Y., Tsumura, K., Arai, T., Senshu, H., Hirata, N., Barucci, M.A., Brunetto, R., Pilorget, C., Poulet, F., Bibring, J.-P., Domingue, D.L., Vilas, F., Takir, D., Palomba, E., Galiano, A., Perna, D., Osawa, T., Komatsu, M., Nakato, A., Arai, T., Takato, N., Matsunaga, T., Arakawa, M., Saiki, T., Wada, K., Kadono, T., Imamura, H., Yano, H., Shirai, K., Hayakawa, M., Okamoto, C., Sawada, H., Ogawa, K., Iijima, Y., Sugita, S., Honda, R., Morota, T., Kameda, S., Tatsumi, E., Cho, Y., Yoshioka, K., Yokota, Y., Sakatani, N., Yamada, M., Kouyama, T., Suzuki, H., Honda, C., Namiki, N., Mizuno, T., Matsumoto, K., Noda, H., Ishihara, Y., Yamada, R., Yamamoto, K., Yoshida, F., Abe, S., Higuchi, A., Yamamoto, Y., Okada, T., Shimaki, Y., Noguchi, R., Miura, A., Hirata, N., Tachibana, S., Yabuta, H., Ishiguro, M., Ikeda, H., Takeuchi, H., Shimada, T., Mori, O., Hosoda, S., Tsukizaki, R., Soldini, S., Ozaki, M., Terui, F., Ogawa, N., Mimasu, Y., Ono, G., Yoshikawa, K., Hirose, C., Fujii, A., Takahashi, T., Kikuchi, S., Takei, Y., Yamaguchi, T., Nakazawa, S., Tanaka, S., Yoshikawa, M., Watanabe, S., Tsuda, Y., 2021. Thermally altered subsurface material of asteroid (162173) Ryugu. Nat Astron 5, 246–250. https://doi.org/10.1038/s41550-020-01271-2

Kurokawa, H., Ehlmann, B.L., De Sanctis, M.C., Lapôtre, M.G.A., Usui, T., Stein, N.T., Prettyman, T.H., Raponi, A., Ciarniello, M., 2020. A Probabilistic Approach to Determination of Ceres' Average Surface Composition From Dawn Visible-Infrared Mapping Spectrometer and Gamma Ray and Neutron Detector Data. Journal of Geophysical Research: Planets 125, e2020JE006606. https://doi.org/10.1029/2020JE006606

Kurokawa, H., Shibuya, T., Sekine, Y., Ehlmann, B.L., Usui, F., Kikuchi, S., Yoda, M., 2022. Distant Formation and Differentiation of Outer Main Belt Asteroids and Carbonaceous Chondrite Parent Bodies. AGU Advances 3, e2021AV000568. https://doi.org/10.1029/2021AV000568

Lauretta, D.S., Balram-Knutson, S.S., Beshore, E., Boynton, W.V., Drouet d'Aubigny, C., DellaGiustina, D.N., Enos, H.L., Golish, D.R., Hergenrother, C.W., Howell, E.S., Bennett, C.A., Morton, E.T., Nolan, M.C., Rizk, B., Roper, H.L., Bartels, A.E., Bos, B.J., Dworkin, J.P., Highsmith, D.E., Lorenz, D.A., Lim, L.F., Mink, R., Moreau, M.C., Nuth, J.A., Reuter, D.C., Simon, A.A., Bierhaus, E.B., Bryan, B.H., Ballouz, R., Barnouin, O.S., Binzel, R.P., Bottke, W.F., Hamilton, V.E., Walsh, K.J., Chesley, S.R., Christensen, P.R., Clark, B.E., Connolly,

H.C., Crombie, M.K., Daly, M.G., Emery, J.P., McCoy, T.J., McMahon, J.W., Scheeres, D.J., Messenger, S., Nakamura-Messenger, K., Righter, K., Sandford, S.A., 2017. OSIRIS-REx: Sample Return from Asteroid (101955) Bennu. Space Sci Rev 212, 925–984. https://doi.org/10.1007/s11214-017-0405-1

Lauretta, D.S., Bartels, A.E., Barucci, M.A., Bierhaus, E.B., Binzel, R.P., Bottke, W.F., Campins, H., Chesley, S.R., Clark, B.C., Clark, B.E., Cloutis, E.A., Connolly, H.C., Crombie, M.K., Delbó, M., Dworkin, J.P., Emery, J.P., Glavin, D.P., Hamilton, V.E., Hergenrother, C.W., Johnson, C.L., Keller, L.P., Michel, P., Nolan, M.C., Sandford, S.A., Scheeres, D.J., Simon, A.A., Sutter, B.M., Vokrouhlický, D., Walsh, K.J., 2015. The OSIRIS-REx target asteroid (101955) Bennu: Constraints on its physical, geological, and dynamical nature from astronomical observations. Meteoritics & Planetary Science 50, 834–849. https://doi.org/10.1111/maps.12353

Matsuoka, M., Nakamura, T., Hiroi, T., Okumura, S., Sasaki, S., 2020. Space Weathering Simulation with Low-energy Laser Irradiation of Murchison CM Chondrite for Reproducing Micrometeoroid Bombardments on C-type Asteroids. ApJL 890, L23. https://doi.org/10.3847/2041-8213/ab72a4

Matsuoka, M., Nakamura, T., Kimura, Y., Hiroi, T., Nakamura, R., Okumura, S., Sasaki, S., 2015. Pulse-laser irradiation experiments of Murchison CM2 chondrite for reproducing space weathering on C-type asteroids. Icarus 254, 135–143. https://doi.org/10.1016/j.icarus.2015.02.029

Matsuoka, M., Nakamura, T., Miyajima, N., Hiroi, T., Imae, N., Yamaguchi, A., 2022. Spectral and mineralogical alteration process of naturally-heated CM and CY chondrites. Geochimica et Cosmochimica Acta 316, 150–167. https://doi.org/10.1016/j.gca.2021.08.042

McAdam, M.M., Sunshine, J.M., Howard, K.T., McCoy, T.M., 2015. Aqueous alteration on asteroids: Linking the mineralogy and spectroscopy of CM and CI chondrites. Icarus 245, 320–332. https://doi.org/10.1016/j.icarus.2014.09.041

McGraw, L.E., Emery, J.P., Thomas, C.A., Rivkin, A.R., Wigton, N.R., McAdam, M., 2022. 3 µm Spectroscopic Survey of Near-Earth Asteroids. Planet. Sci. J. 3, 243. https://doi.org/10.3847/PSJ/ac8ced

McSween, H.Y., 1987. Aqueous alteration in carbonaceous chondrites: Mass balance constraints on matrix mineralogy. Geochimica et Cosmochimica Acta 51, 2469–2477. https://doi.org/10.1016/0016-7037(87)90298-5

McSween, H.Y., Lauretta, D.S., 2006. Meteorites and the Early Solar System II. University of Arizona Press, Tucson.

Milliken, R.E., Hiroi, T., Patterson, W., 2016. THE NASA REFLECTANCE EXPERIMENT LABORATORY (RELAB) FACILITY: PAST, PRESENT, 2.

Milliken, R.E., Mustard, J.F., 2007a. Estimating the water content of hydrated minerals using reflectance spectroscopy: II. Effects of particle size. Icarus 189, 574–588. https://doi.org/10.1016/j.icarus.2006.12.028

Milliken, R.E., Mustard, J.F., 2007b. Estimating the water content of hydrated minerals using reflectance spectroscopy: I. Effects of darkening agents and low-albedo materials. Icarus 189, 550–573. https://doi.org/10.1016/j.icarus.2007.02.017

Miyamoto, M., Zolensky, M.E., 1994. Infrared diffuse reflectance spectra of carbonaceous chondrites: Amount of hydrous minerals. Meteoritics 29, 849–853. https://doi.org/10.1111/j.1945-5100.1994.tb01098.x

Morlok, A., Schiller, B., Weber, I., Daswani, M.M., Stojic, A.N., Reitze, M.P., Gramse, T., Wolters, S.D., Hiesinger, H., Grady, M.M., Helbert, J., 2020. Mid-infrared reflectance spectroscopy of carbonaceous chondrites and Calcium–Aluminum-rich inclusions. Planetary and Space Science 193, 105078. https://doi.org/10.1016/j.pss.2020.105078
Nakamura, E., Kobayashi, K., Tanaka, R., Kunihiro, T., Kitagawa, H., Potiszil, C., Ota, T., Sakaguchi, C., Yamanaka, M., Ratnayake, D.M., Tripathi, H., Kumar, R., Avramescu, M.-L., Tsuchida, H., Yachi, Y., Miura, H., Abe, M., Fukai, R., Furuya, S., Hatakeda, K., Hayashi, T., Hitomi, Y., Kumagai, K., Miyazaki, A., Nakato, A., Nishimura, M., Okada, T., Soejima, H., Sugita, S., Suzuki, A., Usui, T., Yada, T., Yamamoto, D., Yogata, K., Yoshitake, M., Arakawa, M., Fujii, A., Hayakawa, M., Hirata, Naoyuki, Hirata, Naru, Honda, R., Honda, C., Hosoda, S., Iijima, Y., Ikeda, H., Ishiguro, M., Ishihara, Y., Iwata, T., Kawahara, K., Kikuchi, S., Kitazato, K., Matsumoto, K., Matsuoka, M., Michikami, T., Mimasu, Y., Miura, A., Morota, T., Nakazawa, S., Namiki, N., Noda, H., Noguchi, R., Ogawa, N., Ogawa, K., Okamoto, C., Ono, G., Ozaki, M., Saiki, T., Sakatani, N., Sawada, H., Senshu, H., Shimaki, Y., Shirai, K., Takei, Y., Takeuchi, H., Tanaka, S., Tatsumi, E., Terui, F., Tsukizaki, R., Wada, K., Yamada, M., Yamada, T., Yamamoto, Y., Yano, H., Yokota, Y., Yoshihara, K., Yoshikawa, M., Yoshikawa, K., Fujimoto, M., Watanabe, S., Tsuda, Y., 2022. On the origin and evolution of the asteroid Ryugu: A comprehensive geochemical perspective. Proceedings of the Japan Academy, Series B 98, 227–282. https://doi.org/10.2183/pjab.98.015
Nakamura, T., 2005. Post-hydration thermal metamorphism of carbonaceous chondrites. Journal of Mineralogical and Petrological Sciences 100, 260–272. https://doi.org/10.2465/jmps.100.260
Nakamura, T., Matsumoto, M., Amano, K., Enokido, Y., Zolensky, M.E., Mikouchi, T., Genda, H., Tanaka, S., Zolotov, M.Y., Kurosawa, K., Wakita, S., Hyodo, R., Nagano, H., Nakashima, D., Takahashi, Y., Fujioka, Y., Kikuiri, M., Kagawa, E., Matsuoka, M., Brearley, A.J., Tsuchiyama, A., Uesugi, M., Matsuno, J., Kimura, Y., Sato, M., Milliken, R.E., Tatsumi, E., Sugita, S., Hiroi, T., Kitazato, K., Brownlee, D., Joswiak, D.J., Takahashi, M., Ninomiya, K., Takahashi, T., Osawa, T., Terada, K., Brenker, F.E., Tkalcec, B.J., Vincze, L., Brunetto, R., Aléon-Toppani, A., Chan, Q.H.S., Roskosz, M., Viennet, J.-C., Beck, P., Alp, E.E., Michikami, T., Nagaashi, Y., Tsuji, T., Ino, Y., Martinez, J., Han, J., Dolocan, A., Bodnar, R.J., Tanaka, M., Yoshida, H., Sugiyama, K., King, A.J., Fukushi, K., Suga, H., Yamashita, S., Kawai, T., Inoue, K., Nakato, A., Noguchi, T., Vilas, F., Hendrix, A.R., Jaramillo-Correa, C., Domingue, D.L., Dominguez, G., Gainsforth, Z., Engrand, C., Duprat, J., Russell, S.S., Bonato, E., Ma, C., Kawamoto, T., Wada, T., Watanabe, S., Endo, R., Enju, S., Riu, L., Rubino, S., Tack, P., Takeshita, S., Takeichi, Y., Takeuchi, A., Takigawa, A., Takir, D., Tanigaki, T., Taniguchi, A., Tsukamoto, K., Yagi, T., Yamada, S., Yamamoto, K., Yamashita, Y., Yasutake, M., Uesugi, K., Umegaki, I., Chiu, I., Ishizaki, T., Okumura, S., Palomba, E., Pilorget, C., Potin, S.M., Alasli, A., Anada, S., Araki, Y., Sakatani, N., Schultz, C., Sekizawa, O., Sitzman, S.D., Sugiura, K., Sun, M., Dartois, E., De Pauw, E., Dionnet, Z., Djouadi, Z., Falkenberg, G., Fujita, R., Fukuma, T., Gearba, I.R., Hagiya, K., Hu, M.Y., Kato, T., Kawamura, T., Kimura, M., Kubo, M.K., Langenhorst, F., Lantz, C., Lavina, B., Lindner, M., Zhao, J., Vekemans, B., Baklouti, D., Bazi, B., Borondics, F., Nagasawa, S., Nishiyama, G., Nitta, K., Mathurin, J., Matsumoto, T., Mitsukawa, I., Miura, H., Miyake, A., Miyake, Y., Yurimoto, H., Okazaki, R., Yabuta, H., Naraoka, H., Sakamoto, K., Tachibana, S., Connolly, H.C., Lauretta, D.S., Yoshitake, M., Yoshikawa, M., Yoshikawa, K., Yoshihara, K., Yokota, Y., Yogata, K., Yano, H., Yamamoto, Y., Yamamoto, D., Yamada, M., Yamada, T., Yada, T., Wada, K., Usui, T., Tsukizaki, R., Terui, F., Takeuchi, H., Takei, Y., Iwamae, A., Soejima, H., Shirai, K., Shimaki, Y., Senshu, H., Sawada, H., Saiki, T., Ozaki, M., Ono, G., Okada, T., Ogawa, N., Ogawa, K., Noguchi, R., Noda, H., Nishimura, M., Namiki, N., Nakazawa, S., Morota, T., Miyazaki, A., Miura, A., Mimasu, Y., Matsumoto, K., Kumagai, K., Kouyama, T., Kikuchi, S., Kawahara,

K., Kameda, S., Iwata, T., Ishihara, Y., Ishiguro, M., Ikeda, H., Hosoda, S., Honda, R., Honda, C., Hitomi, Y., Hirata, N., Hirata, N., Hayashi, T., Hayakawa, M., Hatakeda, K., Furuya, S., Fukai, R., Fujii, A., Cho, Y., Arakawa, M., Abe, M., Watanabe, S., Tsuda, Y., 2022. Formation and evolution of carbonaceous asteroid Ryugu: Direct evidence from returned samples. Science 379, eabn8671. https://doi.org/10.1126/science.abn8671

Noguchi, T., Matsumoto, T., Miyake, A., Igami, Y., Haruta, M., Saito, H., Hata, S., Seto, Y., Miyahara, M., Tomioka, N., Ishii, H.A., Bradley, J.P., Ohtaki, K.K., Dobrică, E., Leroux, H., Le Guillou, C., Jacob, D., de la Peña, F., Laforet, S., Marinova, M., Langenhorst, F., Harries, D., Beck, P., Phan, T.H.V., Rebois, R., Abreu, N.M., Gray, J., Zega, T., Zanetta, P.-M., Thompson, M.S., Stroud, R., Burgess, K., Cymes, B.A., Bridges, J.C., Hicks, L., Lee, M.R., Daly, L., Bland, P.A., Zolensky, M.E., Frank, D.R., Martinez, J., Tsuchiyama, A., Yasutake, M., Matsuno, J., Okumura, S., Mitsukawa, I., Uesugi, K., Uesugi, M., Takeuchi, A., Sun, M., Enju, S., Takigawa, A., Michikami, T., Nakamura, T., Matsumoto, M., Nakauchi, Y., Abe, M., Arakawa, M., Fujii, A., Hayakawa, M., Hirata, Naru, Hirata, Naoyuki, Honda, R., Honda, C., Hosoda, S., Iijima, Y., Ikeda, H., Ishiguro, M., Ishihara, Y., Iwata, T., Kawahara, K., Kikuchi, S., Kitazato, K., Matsumoto, K., Matsuoka, M., Mimasu, Y., Miura, A., Morota, T., Nakazawa, S., Namiki, N., Noda, H., Noguchi, R., Ogawa, N., Ogawa, K., Okada, T., Okamoto, C., Ono, G., Ozaki, M., Saiki, T., Sakatani, N., Sawada, H., Senshu, H., Shimaki, Y., Shirai, K., Sugita, S., Takei, Y., Takeuchi, H., Tanaka, S., Tatsumi, E., Terui, F., Tsukizaki, R., Wada, K., Yamada, M., Yamada, T., Yamamoto, Y., Yano, H., Yokota, Y., Yoshihara, K., Yoshikawa, M., Yoshikawa, K., Fukai, R., Furuya, S., Hatakeda, K., Hayashi, T., Hitomi, Y., Kumagai, K., Miyazaki, A., Nakato, A., Nishimura, M., Soejima, H., Suzuki, A.I., Usui, T., Yada, T., Yamamoto, D., Yogata, K., Yoshitake, M., Connolly, H.C., Lauretta, D.S., Yurimoto, H., Nagashima, K., Kawasaki, N., Sakamoto, N., Okazaki, R., Yabuta, H., Naraoka, H., Sakamoto, K., Tachibana, S., Watanabe, S., Tsuda, Y., 2023. A dehydrated space-weathered skin cloaking the hydrated interior of Ryugu. Nat Astron 7, 170–181. https://doi.org/10.1038/s41550-022-01841-6

Pilorget, C., Okada, T., Hamm, V., Brunetto, R., Yada, T., Loizeau, D., Riu, L., Usui, T., Moussi-Soffys, A., Hatakeda, K., Nakato, A., Yogata, K., Abe, M., Aléon-Toppani, A., Carter, J., Chaigneau, M., Crane, B., Gondet, B., Kumagai, K., Langevin, Y., Lantz, C., Le Pivert-Jolivet, T., Lequertier, G., Lourit, L., Miyazaki, A., Nishimura, M., Poulet, F., Arakawa, M., Hirata, N., Kitazato, K., Nakazawa, S., Namiki, N., Saiki, T., Sugita, S., Tachibana, S., Tanaka, S., Yoshikawa, M., Tsuda, Y., Watanabe, S., Bibring, J.-P., 2022. First compositional analysis of Ryugu samples by the MicrOmega hyperspectral microscope. Nat Astron 6, 221–225. https://doi.org/10.1038/s41550-021-01549-z

Potin, S., Beck, P., Usui, F., Bonal, L., Vernazza, P., Schmitt, B., 2020. Style and intensity of hydration among C-complex asteroids: A comparison to desiccated carbonaceous chondrites. Icarus 348, 113826. https://doi.org/10.1016/j.icarus.2020.113826

Potin, S.M., Douté, S., Kugler, B., Forbes, F., 2022. The impact of asteroid shapes and topographies on their reflectance spectroscopy. Icarus 376, 114806. https://doi.org/10.1016/j.icarus.2021.114806

Prettyman, T.H., Yamashita, N., Toplis, M.J., McSween, H.Y., Schörghofer, N., Marchi, S., Feldman, W.C., Castillo-Rogez, J., Forni, O., Lawrence, D.J., Ammannito, E., Ehlmann, B.L., Sizemore, H.G., Joy, S.P., Polanskey, C.A., Rayman, M.D., Raymond, C.A., Russell, C.T., 2017. Extensive water ice within Ceres' aqueously altered regolith: Evidence from nuclear spectroscopy. Science 355, 55–59. https://doi.org/10.1126/science.aah6765

Rivkin, A.S., Howell, E.S., Emery, J.P., Volquardsen, E.L., DeMeo, F.E., 2012. Toward a taxonomy of asteroid spectra in the 3-µm region. Presented at the European Planetary Science Congress 2012.

Rubin, A.E., Trigo-Rodríguez, J.M., Huber, H., Wasson, J.T., 2007. Progressive aqueous alteration of CM carbonaceous chondrites. Geochimica et Cosmochimica Acta 71, 2361–2382. https://doi.org/10.1016/j.gca.2007.02.008

Russell, C.T., Raymond, C.A., 2011. The Dawn Mission to Vesta and Ceres. Space Sci Rev 163, 3–23. https://doi.org/10.1007/s11214-011-9836-2

Schultz, C., Anzures, B.A., Milliken, R.E., Hiroi, T., Robertson, K., 2023. Assessing the spatial variability of the 3 µm OH/H2O absorption feature in CM2 carbonaceous chondrites. Meteoritics & Planetary Science 58, 170–194. https://doi.org/10.1111/maps.13946

Takir, D., Emery, J.P., 2012. Outer Main Belt asteroids: Identification and distribution of four 3-µm spectral groups. Icarus 219, 641–654. https://doi.org/10.1016/j.icarus.2012.02.022

Takir, D., Emery, J.P., McSween, H.Y., 2015. Toward an understanding of phyllosilicate mineralogy in the outer main asteroid belt. Icarus 257, 185–193. https://doi.org/10.1016/j.icarus.2015.04.042

Takir, D., Emery, J.P., Mcsween, H.Y., Hibbitts, C.A., Clark, R.N., Pearson, N., Wang, A., 2013. Nature and degree of aqueous alteration in CM and CI carbonaceous chondrites. Meteorit Planet Sci 20. https://doi.org/10.1111/maps.12171

Takir, D., Neumann, W., Raymond, S.N., Emery, J.P., Trieloff, M., 2023. Late accretion of Ceres-like asteroids and their implantation into the outer main belt. Nat Astron 7, 524–533. https://doi.org/10.1038/s41550-023-01898-x

Takir, D., Reddy, V., Sanchez, J.A., Shepard, M.K., Emery, J.P., 2016. Detection of water and/or hydroxyl on asteroid (16) Psyche. AJ 153, 31. https://doi.org/10.3847/1538-3881/153/1/31

Vernazza, P., Delbo, M., King, P.L., Izawa, M.R.M., Olofsson, J., Lamy, P., Cipriani, F., Binzel, R.P., Marchis, F., Merín, B., Tamanai, A., 2012. High surface porosity as the origin of emissivity features in asteroid spectra. Icarus 221, 1162–1172. https://doi.org/10.1016/j.icarus.2012.04.003

Watanabe, S., Tsuda, Y., Yoshikawa, M., Tanaka, S., Saiki, T., Nakazawa, S., 2017. Hayabusa2 Mission Overview. Space Sci Rev 208, 3–16. https://doi.org/10.1007/s11214-017-0377-1

Weidenschilling, S.J., 2019. Accretion of the asteroids: Implications for their thermal evolution. Meteoritics & Planetary Science 54, 1115–1132. https://doi.org/10.1111/maps.13270

Yokoyama, T., Nagashima, K., Nakai, I., Young, E.D., Abe, Y., Aléon, J., Alexander, C.M.O., Amari, S., Amelin, Y., Bajo, K., Bizzarro, M., Bouvier, A., Carlson, R.W., Chaussidon, M., Choi, B.-G., Dauphas, N., Davis, A.M., Rocco, T.D., Fujiya, W., Fukai, R., Gautam, I., Haba, M.K., Hibiya, Y., Hidaka, H., Homma, H., Hoppe, P., Huss, G.R., Ichida, K., Iizuka, T., Ireland, T.R., Ishikawa, A., Ito, M., Itoh, S., Kawasaki, N., Kita, N.T., Kitajima, K., Kleine, T., Komatani, S., Krot, A.N., Liu, M.-C., Masuda, Y., McKeegan, K.D., Morita, M., Motomura, K., Moynier, F., Nguyen, A., Nittler, L., Onose, M., Pack, A., Park, C., Piani, L., Qin, L., Russell, S.S., Sakamoto, N., Schönbächler, M., Tafla, L., Tang, H., Terada, K., Terada, Y., Usui, T., Wada, S., Wadhwa, M., Walker, R.J., Yamashita, K., Yin, Q.-Z., Yoneda, S., Yui, H., Zhang, A.-C., ConnollyJr, H.C., Lauretta, D.S., Nakamura, T., Naraoka, H., Noguchi, T., Okazaki, R., Sakamoto, K., Yabuta, H., Abe, M., Arakawa, M., Fujii, A., Hayakawa, M., Hirata, Naoyuki, Hirata, Naru, Honda, R., Honda, C., Hosoda, S., Iijima, Y., Ikeda, H., Ishiguro, M., Ishihara, Y., Iwata, T., Kawahara, K., Kikuchi, S., Kitazato, K., Matsumoto, K., Matsuoka, M., Michikami, T., Mimasu, Y., Miura, A., Morota, T., Nakazawa, S., Namiki, N., Noda, H., Noguchi, R., Ogawa, N., Ogawa, K., Okada, T., Okamoto, C., Ono, G., Ozaki, M., Saiki, T.,

Sakatani, N., Sawada, H., Senshu, H., Shimaki, Y., Shirai, K., Sugita, S., Takei, Y., Takeuchi, H., Tanaka, S., Tatsumi, E., Terui, F., Tsuda, Y., Tsukizaki, R., Wada, K., Watanabe, S., Yamada, M., Yamada, T., Yamamoto, Y., Yano, H., Yokota, Y., Yoshihara, K., Yoshikawa, M., Yoshikawa, K., Furuya, S., Hatakeda, K., Hayashi, T., Hitomi, Y., Kumagai, K., Miyazaki, A., Nakato, A., Nishimura, M., Soejima, H., Suzuki, A., Yada, T., Yamamoto, D., Yogata, K., Yoshitake, M., Tachibana, S., Yurimoto, H., 2022. Samples returned from the asteroid Ryugu are similar to Ivuna-type carbonaceous meteorites. Science. https://doi.org/10.1126/science.abn7850

Yumoto, K., Tatsumi, E., Kouyama, T., Golish, D.R., Kameda, S., Sato, H., Rizk, B., DellaGiustina, D.N., Yokota, Y., Suzuki, H., de León, J., Campins, H., Licandro, J., Popescu, M., Rizos, J.L., Honda, R., Yamada, M., Morota, T., Sakatani, N., Cho, Y., Honda, C., Matsuoka, M., Hayakawa, M., Sawada, H., Ogawa, K., Yamamoto, Y., Sugita, S., Lauretta, D.S., 2023. Cross calibration between Hayabusa2/ONC-T and OSIRIS-REx/MapCam for comparative analyses between asteroids Ryugu and Bennu. https://doi.org/10.48550/arXiv.2306.13321

Zolensky, M., Barrett, R., Browning, L., 1993. Mineralogy and composition of matrix and chondrule rims in carbonaceous chondrites. Geochimica et Cosmochimica Acta 57, 3123–3148. https://doi.org/10.1016/0016-7037(93)90298-B

# Supplementary materials

Tab. S1. List of studied carbonaceous chondrite samples ID and spectrum ID from NASA RELAB spectral database (http://www.planetary.brown.edu/relabdata/). For QUE97077, Murchison, Murray, Nogoya, Mighei and Cold Bokkeveld meteorite, spectral data currently unavailable at RELAB database, therefore we have uploaded the spectral data in the attached file '6 CM2 SPECTRAL'.

| Meteorite name | Classification | Specimen ID | Measurement ID |
|---|---|---|---|
| Alais | CI1 | MT-KTH-264 | BIR1MT264 |
| Orgueil | CI1 | MT-JMS-191 | BMR1MT191 |
| Ivuna | CI1 | MX-TXH-064 | BIR2MX064P |
| Moapa Valley | CM1 | MT-KTH-267 | BIR1MT267 |
| Al Rais | CR2 | MT-KTH-265 | BIR1MT265 |
| ALHA77307 | CO3 | MT-KTH-283 | BIR1MT283 |
| Allende | CV3 | MT-KTH-268 | BIR1MT268 |
| EET92002 | CK4 | MC-RPB-003 | N1MC03 |
| Y-693 | CK4 | MB-TXH-077 | MMMB77 |
| EET87507 | CK4/5 | MB-TXH-092 | MMMB92 |
| Tagish Lake | C2ung | MT-MEZ-318 | BMR1MT318A |

Tab. S2. List of studied CM2 Murchison heating experiment samples ID and spectrum ID from RELAB.

| Specimen Description | Specimen ID | Measurement ID |
|---|---|---|
| Murchison unheated | MB-TXH-064-D4 | BIR1MB064D4 |
| Murchison heated at 400°C | MB-TXH-064-E4 | BIR1MB064E4 |
| Murchison heated at 500°C | MB-TXH-064-F4 | BIR1MB064F4 |
| Murchison heated at 600°C | MB-TXH-064-G4 | BIR1MB064G4 |
| Murchison heated at 700°C | MB-TXH-064-H4 | BIR1MB064H4 |
| Murchison heated at 800°C | MB-TXH-064-I4 | BIR1MB064I4 |
| Murchison heated at 900°C | MB-TXH-064-J4 | BIR1MB064J4 |
| Murchison heated at 1000°C | MB-TXH-064-K4 | BIR1MB064K4 |

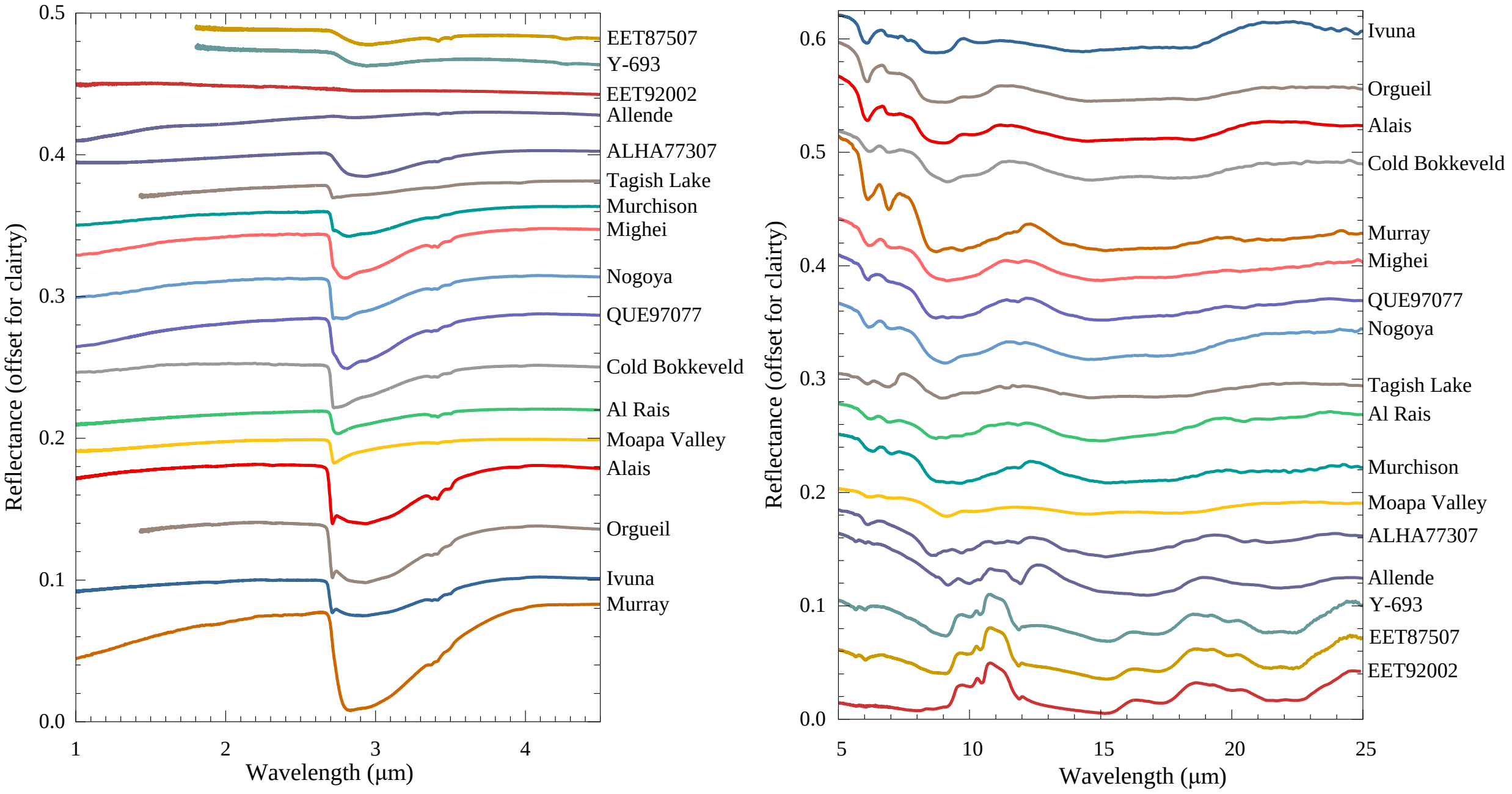

Fig. S1. Reflectance spectra of 17 carbonaceous chondrites in the 1–25 μm spectral range.

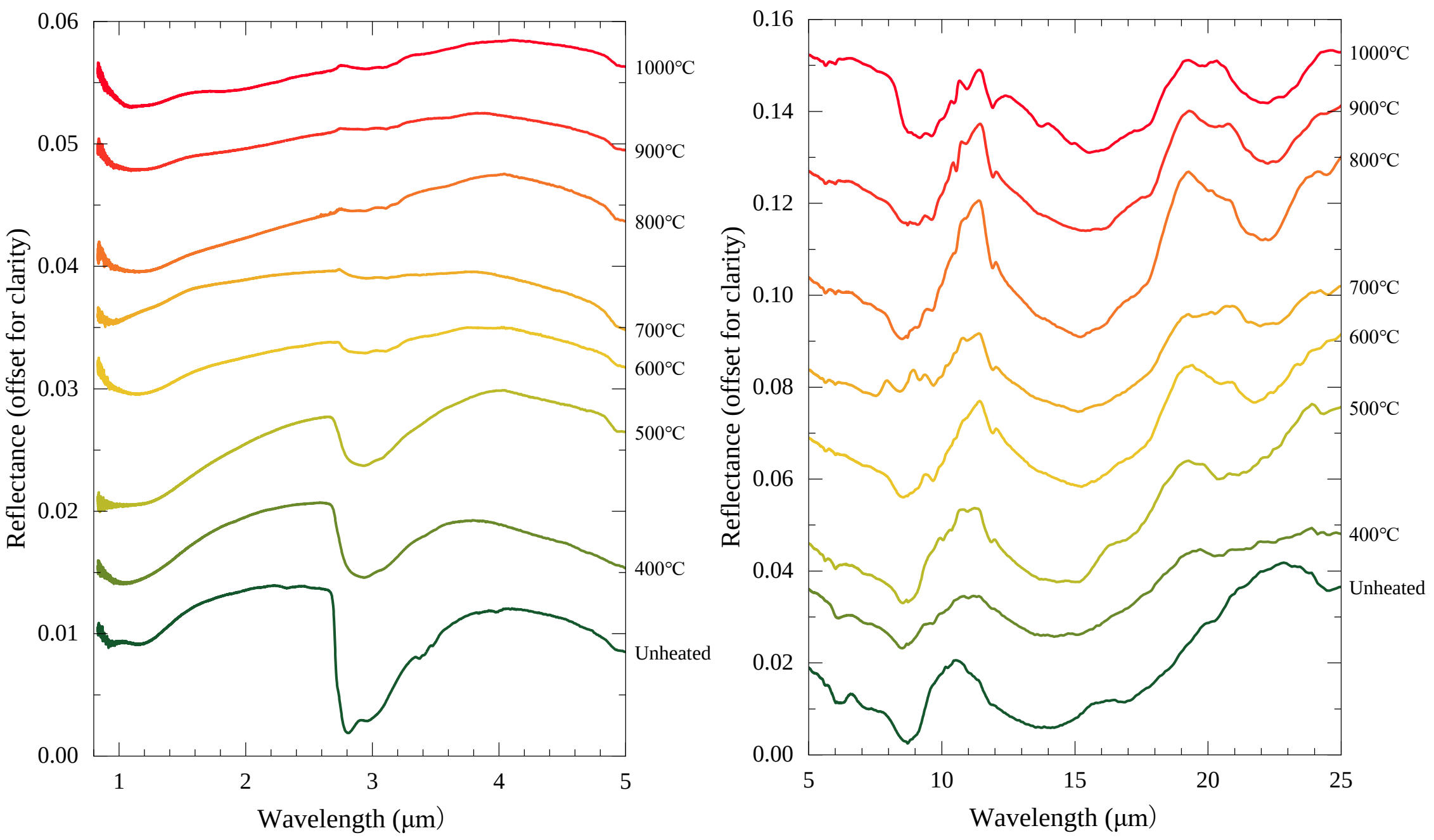

Fig. S2. Reflectance spectra of Murchison sample after heating at different temperatures at 0.8–25 μm spectral range.

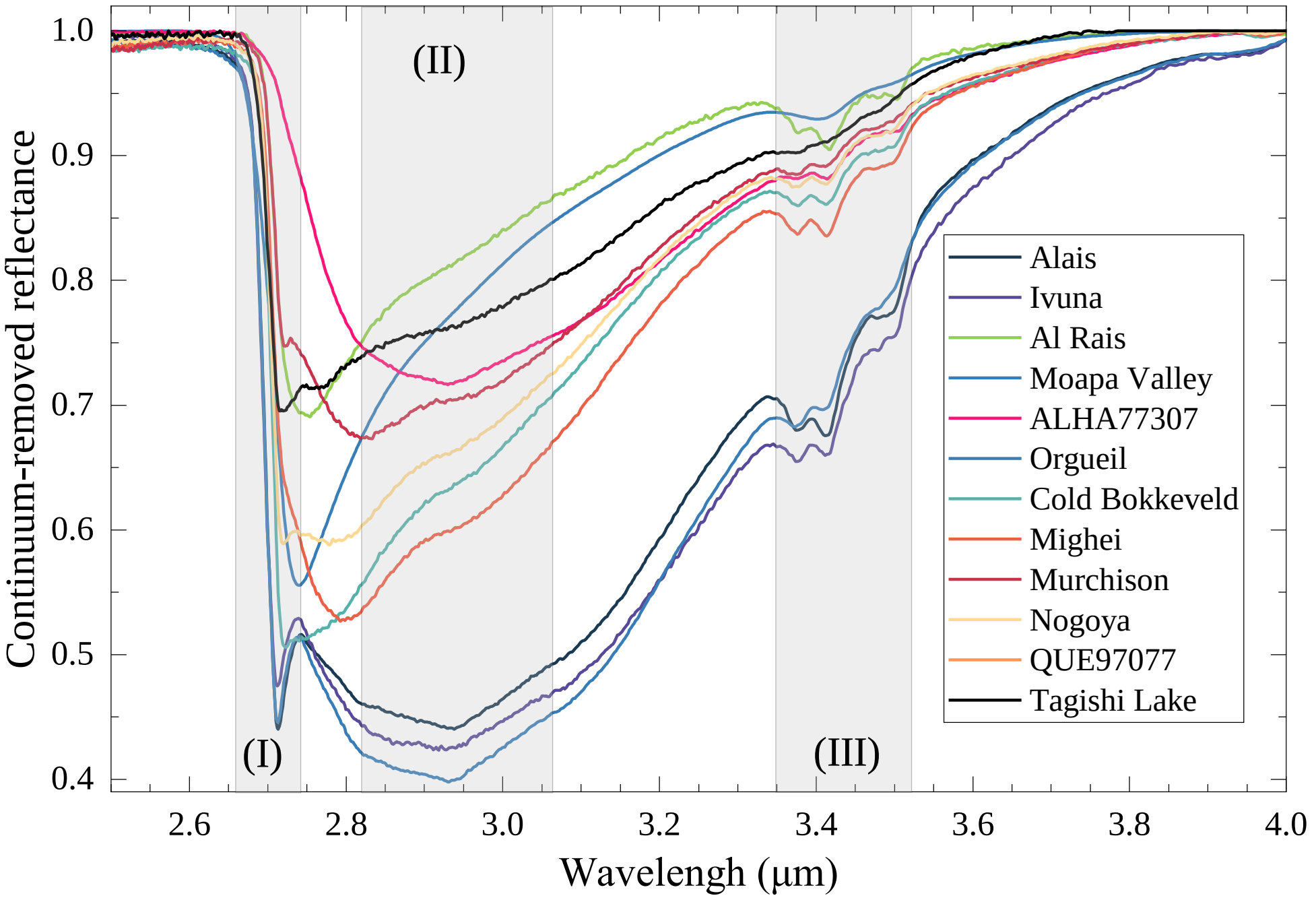

Fig. S3. Continuum-removed specta of aqueous altered carbonaceous chondrites (for more details, please see the sample description in section 2.1 of the main text) at 3 μm band, the shaded areas in the figure indicate: (I) Mg-rich serpentine-like minerals, (II) Fe-rich cronstedtite-like minerals, and $H_2O$, (III) aliphatic and aromatic organic compounds. Note that absorption features around 2.9 μm are mainly from terrestrial adsorbed water, and are most pronounced in the CI samples.

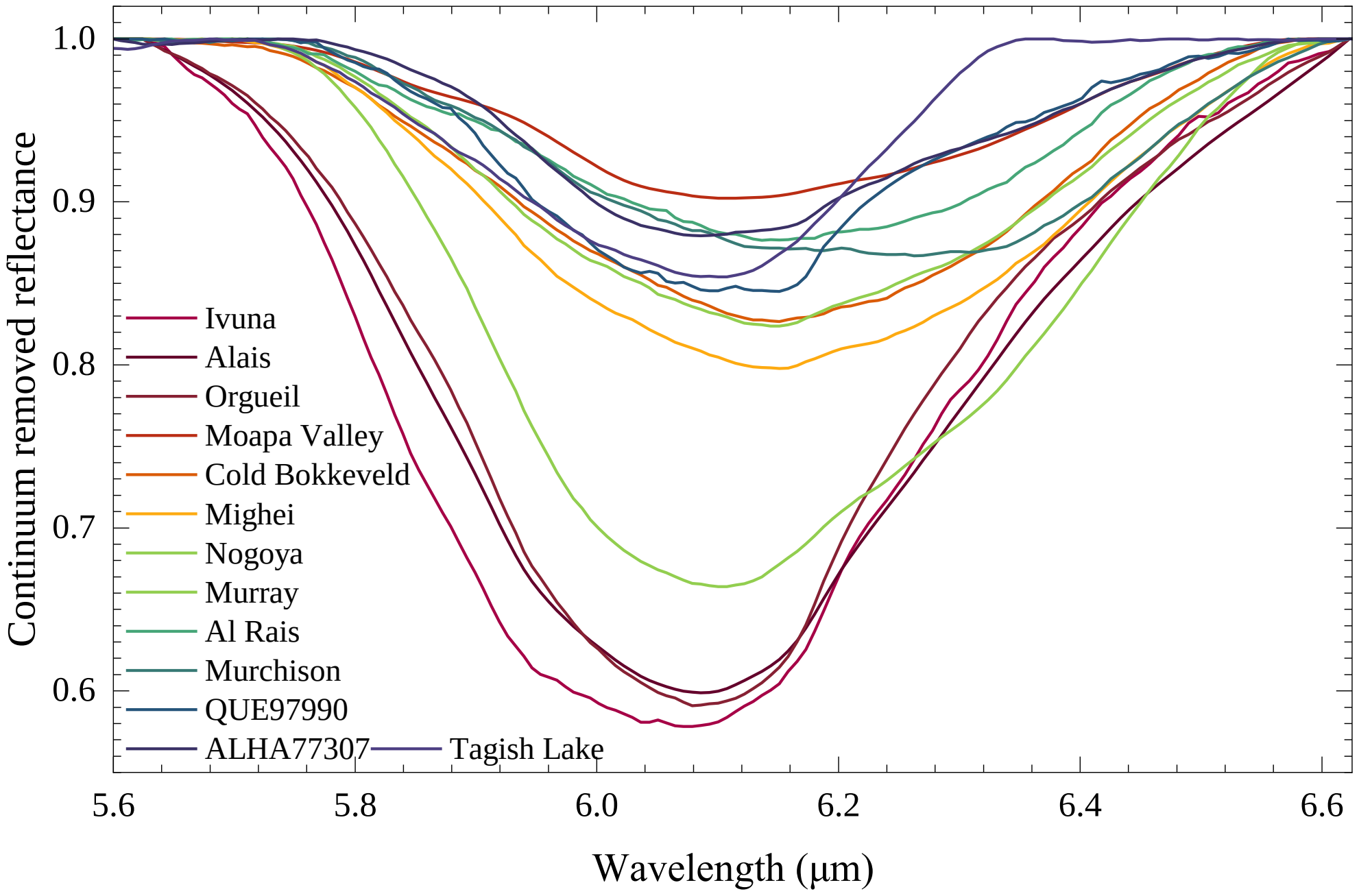

Fig. S4. Continuum-removed spectra of aqueous altered carbonaceous chondrites of the 6 μm band, note that absorption of the CO3 ALHA77307 meteorite is mainly from terrestrial adsorbed water.